\documentclass[12pt,preprint]{aastex63}

\usepackage[T1]{tipa}
\usepackage{amsmath}
\usepackage{amssymb}
\usepackage{color}
\usepackage{graphicx}
\usepackage{hyperref}
\usepackage{natbib}
\usepackage{savesym}
\savesymbol{tablenum}
\usepackage{multirow}
\usepackage{siunitx}
\usepackage{threeparttablex}
\usepackage{booktabs}
\usepackage{marginnote}
\usepackage{changepage,threeparttable}
\usepackage{blkarray}
\usepackage{makecell}
\usepackage{wrapfig}
\restoresymbol{SIX}{tablenum}

\DeclareTextFontCommand{\textipa}{%
  \fontfamily{cmss}\tipaencoding
}

\newcommand{\DM}[1]{\ensuremath{\text{DM}_\text{#1}}}
\newcommand{\barDM}[1]{\ensuremath{\overline{\text{DM}}_\text{#1}}}
\newcommand{\dmu}{pc~cm$^{-3}$}
\newcommand{\msun}{M$_\odot$}
\newcommand{\kkoname}{k'ni\textipa{P}atn k'l$\left._\mathrm{\smile}\right.$stk'masqt}

\begin{document}

\title{Constraining Gas Mass Fractions in Galaxy Groups and Clusters with the First CHIME/FRB Outrigger}

\shorttitle{Probing galaxy groups with FRBs}
\shortauthors{CHIME/FRB Collaboration: A.~Lanman, \emph{et al.}}

%%%%%%%%%%%%%%%%% BEGIN AUTHOR SCRIPT PASTE %%%%%%%%%%%%%%%%%%%%%%
% do not edit this file!  It is auto-generated from the authorship spreadsheet.
\author[0000-0003-2116-3573]{Adam E.~Lanman}
  \affiliation{MIT Kavli Institute for Astrophysics and Space Research, Massachusetts Institute of Technology, 77 Massachusetts Ave, Cambridge, MA 02139, USA}
  \affiliation{Department of Physics, Massachusetts Institute of Technology, 77 Massachusetts Ave, Cambridge, MA 02139, USA}
\author[0000-0003-3801-1496]{Sunil Simha}
  \affiliation{Department of Astronomy and Astrophysics, University of Chicago, William Eckhardt Research Center, 5640 South Ellis Avenue, Chicago, IL 60637, USA}
  \affiliation{Center for Interdisciplinary Research in Astrophysics, Northwestern University, 1800 Sherman Ave., Evanston, IL 60201, USA}
\author[0000-0002-4279-6946]{Kiyoshi W.~Masui}
  \affiliation{MIT Kavli Institute for Astrophysics and Space Research, Massachusetts Institute of Technology, 77 Massachusetts Ave, Cambridge, MA 02139, USA}
  \affiliation{Department of Physics, Massachusetts Institute of Technology, 77 Massachusetts Ave, Cambridge, MA 02139, USA}
\author[0000-0002-7738-6875]{J.~Xavier Prochaska}
  \affiliation{Department of Astronomy and Astrophysics, University of California, Santa Cruz, 1156 High Street, Santa Cruz, CA 95064, USA}
\author[0000-0001-7674-5066]{Rachel Darlinger}
  \affiliation{Department of Physics, McGill University, 3600 rue University, Montr\'eal, QC H3A 2T8, Canada}
  \affiliation{Trottier Space Institute at McGill University, 3550 rue University, Montr\'eal, QC H3A 2A7, Canada}
\author[0000-0003-4098-5222]{Fengqiu Adam Dong}
  \affiliation{National Radio Astronomy Observatory, 520 Edgemont Rd, Charlottesville, VA 22903, USA}
\author[0000-0002-3382-9558]{B.~M.~Gaensler}
  \affiliation{Department of Astronomy and Astrophysics, University of California, Santa Cruz, 1156 High Street, Santa Cruz, CA 95064, USA}
  \affiliation{Dunlap Institute for Astronomy \& Astrophysics, University of Toronto, 50 St.~George Street, Toronto, ON M5S 3H4, Canada}
  \affiliation{David A.~Dunlap Department of Astronomy \& Astrophysics, University of Toronto, 50 St.~George Street, Toronto, ON M5S 3H4, Canada}
\author[0000-0003-3457-4670]{Ronniy C.~Joseph}
  \affiliation{Department of Physics, McGill University, 3600 rue University, Montr\'eal, QC H3A 2T8, Canada}
  \affiliation{Trottier Space Institute at McGill University, 3550 rue University, Montr\'eal, QC H3A 2A7, Canada}
\author[0000-0003-4810-7803]{Jane Kaczmarek}
  \affiliation{Dominion Radio Astrophysical Observatory, Herzberg Research Centre for Astronomy and Astrophysics, National Research Council Canada, PO Box 248, Penticton, BC V2A 6J9, Canada}
  \affiliation{SKA Observatory, Science Operations Centre, 26 Dick Perry Avenue. Kensington WA 6151}
\author[0009-0007-5296-4046]{Lordrick Kahinga}
  \affiliation{Department of Astronomy and Astrophysics, University of California, Santa Cruz, 1156 High Street, Santa Cruz, CA 95064, USA}
  \affiliation{Department of Physics, College of Natural and Mathematical Sciences, University of Dodoma, 1 Benjamin Mkapa Road, 41218 Iyumbu, Dodoma 259, Tanzania}
\author[0009-0004-4176-0062]{Afrokk Khan}
  \affiliation{Department of Physics, McGill University, 3600 rue University, Montr\'eal, QC H3A 2T8, Canada}
  \affiliation{Trottier Space Institute at McGill University, 3550 rue University, Montr\'eal, QC H3A 2A7, Canada}
\author[0000-0002-4209-7408]{Calvin Leung}
  \affiliation{Department of Astronomy, University of California, Berkeley, CA 94720, USA}
  \affiliation{Miller Institute for Basic Research, University of California, Berkeley, CA 94720, USA}
\author[0000-0003-4584-8841]{Lluis Mas-Ribas}
  \affiliation{Department of Astronomy and Astrophysics, University of California, Santa Cruz, 1156 High Street, Santa Cruz, CA 95064, USA}
\author[0009-0008-7264-1778]{Swarali Shivraj Patil}
  \affiliation{Department of Physics and Astronomy, West Virginia University, PO Box 6315, Morgantown, WV 26506, USA}
  \affiliation{Center for Gravitational Waves and Cosmology, West Virginia University, Chestnut Ridge Research Building, Morgantown, WV 26505, USA}
\author[0000-0002-8912-0732]{Aaron~B.~Pearlman}
  \altaffiliation{NASA Hubble Fellow.}
  \affiliation{MIT Kavli Institute for Astrophysics and Space Research, Massachusetts Institute of Technology, 77 Massachusetts Ave, Cambridge, MA 02139, USA}
  \affiliation{Department of Physics, Massachusetts Institute of Technology, 77 Massachusetts Ave, Cambridge, MA 02139, USA}
  \affiliation{Department of Physics, McGill University, 3600 rue University, Montr\'eal, QC H3A 2T8, Canada}
  \affiliation{Trottier Space Institute, McGill University, 3550 rue University, Montr\'eal, QC H3A 2A7, Canada}
\author[0000-0002-4623-5329]{Mawson Sammons}
  \affiliation{Department of Physics, McGill University, 3600 rue University, Montr\'eal, QC H3A 2T8, Canada}
  \affiliation{Trottier Space Institute at McGill University, 3550 rue University, Montr\'eal, QC H3A 2A7, Canada}
\author[0000-0002-6823-2073]{Kaitlyn Shin}
  \affiliation{Cahill Center for Astronomy and Astrophysics, MC 249-17 California Institute of Technology, Pasadena CA 91125, USA}
\author[0000-0002-2088-3125]{Kendrick Smith}
  \affiliation{Perimeter Institute of Theoretical Physics, 31 Caroline Street North, Waterloo, ON N2L 2Y5, Canada}
\author[0000-0002-1491-3738]{Haochen Wang}
  \affiliation{MIT Kavli Institute for Astrophysics and Space Research, Massachusetts Institute of Technology, 77 Massachusetts Ave, Cambridge, MA 02139, USA}
  \affiliation{Department of Physics, Massachusetts Institute of Technology, 77 Massachusetts Ave, Cambridge, MA 02139, USA}
% Unique acks:
\newcommand{\allacks}{
A.B.P. acknowledges support by NASA through the NASA Hubble Fellowship grant HST-HF2-51584.001-A awarded by the Space Telescope Science Institute, which is operated by the Association of Universities for Research in Astronomy, Inc., under NASA contract NAS5-26555. A.B.P. also acknowledges prior support from a Banting Fellowship, a McGill Space Institute~(MSI) Fellowship, and a Fonds de Recherche du Quebec -- Nature et Technologies~(FRQNT) Postdoctoral Fellowship.
C. L. acknowledges support from the Miller Institute for Basic Research at UC Berkeley.
F.A.D is a NRAO Jansky Fellow; The National Radio Astronomy Observatory and Green Bank Observatory are facilities of the U.S. National Science Foundation operated under cooperative agreement by Associated Universities, Inc.
K.W.M. holds the Adam J. Burgasser Chair in Astrophysics and recieved support from NSF grant 2018490.
M.W.S. acknowledges support from the Trottier Space Institute Fellowship program.
R.D. is supported by the Canada Excellence Research Chair in Transient Astrophysics (CERC-2022-00009)
S.S.P. is supported by the National Science Foundation under grant AST-2407399.
}

%%%%%%%%%%%%%%%% END AUTHOR SCRIPT PASTE %%%%%%%%%%%%%%

\correspondingauthor{Adam E. Lanman}
\email{alanman@mit.edu}

\begin{abstract}
In recent years, localized fast radio bursts (FRBs) have emerged as a powerful tool to study the structure of the baryonic matter in the universe. Their dispersion measures (DMs) scale linearly with electron density independent of gas temperature, making them particularly well suited to studying the intragroup medium (IGrM), where traditional probes such as X-ray emission and the SZ effect are weak. Evidence suggests that the gas in group mass halos ($M_{500}\sim10^{13}$--$10^{14}$~\msun{}) is strongly affected by galactic feedback, causing  deviations from cluster scaling relations. Three FRBs from the first CHIME/FRB Outrigger sample come from host galaxies found within or behind galaxy clusters and groups. We estimate the DM contribution of each ICM/IGrM by integrating different halo density profiles, accounting for uncertainties in halo mass and the host galaxy line of sight distance. For the more massive halos, predicted cluster DMs agree with the extragalactic DM budget. One burst, FRB 20230703A, intersects three groups yet has a low extragalactic DM. By comparing model predictions with the measured DM, we constrain the gas mass fraction $f_g(R)$ in these halos. Comparing with published $M$--$f_g$ relations, we find consistency with recent eROSITA results at $R_{500}$ and mild tension at $R_{200}$ and with earlier X-ray–based relations. As CHIME/FRB Outriggers build a large catalog of localized FRBs, many additional sightlines through groups and clusters will be obtained. These will enable systematic tests of intragroup and intracluster gas properties and sharpen constraints on the distribution of baryons in massive halos.
\end{abstract}

%\tableofcontents

\section{Introduction}

Galaxy clusters, with masses $\sim 10^{14} $--$ 10^{15}$~\msun{}, are the largest gravitationally-bound structures in the universe and are exceptionally useful tools in cosmology and astrophysics. The evolution of massive clusters is dominated by their gravity, leading them to have thermodynamic properties that depend primarily on mass and redshift, and are independent of formation history \citep{Kaiser:1986}. The bulk of baryons in clusters exists in a hot ($k T_x \sim 2 $--$ 4$~keV), centralized \emph{intracluster medium} (ICM), which is typically observable in X-ray emission and through the thermal Sunyaev-Zeldovich effect (tSZ). Rich galaxy populations in clusters, as well as the resolvable X-ray emitting ICM, allow for robust measurements of the total halo mass \citep{Pratt:2010}. Being at the higher end of the halo mass function, cluster mass measurements are highly constraining on the cosmological parameters $\Omega_m$ (the total mass density) and $\sigma_8$ (the matter power spectrum amplitude at 8~$h^{-1}$~Mpc) \citep{Allen:2011, Kravtsov:Borgani:2012}.

Given their importance in cosmology, understanding how baryons are distributed within clusters is essential. One quantity of cosmological interest is the gas mass fraction within the halo, relative to the cosmic baryon fraction $f_B = \Omega_b/\Omega_m$:
\begin{equation}
	f_g(r) = \frac{M_g(r)}{f_B M(r)},
	\label{eq:fg_def}
\end{equation}
where $M_g(r)$ is the mass of ionized gas (i.e., not bound into stars) and $M(r)$ is the total mass contained within radius $r$. The gas fraction is most commonly quoted as $f_{g,500} \equiv f_g(R_{500})$, as typical X-ray observations are only sensitive out to $R_{500}$ (the radius within which the mean matter density is 500$\times$ the critical density). Due to their strong binding energy compared with astrophysical processes within them, clusters are expected to retain nearly all of their baryons from formation, such that $f_{g}(R_v) \approx 1$ at the virial radius $R_v$.\footnote{In this paper, we will commonly refer to three different virial radius conventions, denoted -- $R_{500} < R_{200} < R_v$, all of which are defined in App.~\ref{app:virial_conv}.} Measurements of $f_g$ at smaller radii, despite not capturing the full extent of the halo, have been successfully used as complementary probes of various cosmological parameters. Under the assumption that $f_g$ is largely independent of redshift for massive, relaxed clusters, its scaling with angular diameter distance depends on $\Omega_m$ and $\Omega_\Lambda$ \citep{Mantz:Morris:2021,Ettori:2009,Sasaki:1996,Pen:1997}. In combination with other probes, the gas fraction has also been used to constrain deviations in the fine structure constant \citep{Ferreira:Holanda:2024}, and can potentially break degeneracies among dark sector models \citep{Barbosa:Marttens:2024}.

Lower-mass \emph{galaxy groups} ($\sim 10^{13}$~\msun{}) are an important piece of the picture, bridging the gap between massive galaxy clusters and individual galaxies. Since structure formation proceeds hierarchically, with lower-mass groups forming first and merging into more massive structures, group evolution is key to understanding the formation of clusters, as well as how environments of galaxies may influence their development. Furthermore, $10^{13.5}$~\msun{} halos outnumber $10^{14.5}$ by roughly a factor of 20 \citep{Murray:Power:Rowbotham:2013}, allowing for more powerful statistics for precision cosmology. However, being relatively smaller in scale and binding energy, the macroscopic properties of galaxy groups are more influenced by the non-gravitational behavior of member galaxies -- particularly AGN and stellar feedback. A consequence of this feedback is that the gas fraction $f_{g,500}$, is significantly lower than seen in higher-mass halos. This has been challenging to model in hydrodynamic simulations, as excessive feedback can quench star formation in member galaxies. \cite{Eckert:2021} compiles a range of simulated values for $f_{g,500}$, showing that despite recent simulations being calibrated to observed stellar masses, different feedback prescriptions make strongly different predictions for IGrM baryon content.

\cite{Popesso:Biviano:2024} (hereafter, P24) estimated $f_g$ for a range of halo masses using stacked data from eROSITA, providing both new estimates of $f_{g,500}$ and the first ever estimates of $f_{g,200}$ for galaxy groups, even extending down to $M_{500} \sim 10^{12.5}$ halos. They find power-law fits to the $f_g $~\textendash~$ M_\text{halo}$ relations in decent agreement with prior fits, but do find that $f_g$ of low-mass halos are less than half that predicted by previous studies. They interpret this as a selection effect -- previous studies only detected the more X-ray luminous group halos, which may be more skewed to higher $f_g$ -- highlighting the importance of more sensitive X-ray surveys like eROSITA.

The IGrM of lower-mass groups has been observed in metal absorption \citep{Nielsen:Kacprzak:2018,Stocke:Keeney:2019} and emission \citep{Epinat:Contini:2024}, as well as in 21 cm \citep{Koribalski:2020, Borthakur:Yun:2010, Pisano:Wakker:2004}, revealing a complex, dynamic environment co-evolving with member galaxies. A more recent probe made possible by high-resolution radio surveying instruments is the RM grid technique \citep{Beck:Gaensler:2004}, which has detected the presence of a magnetized IGrM through the rotation measures of unresolved background radio sources \citep{Anderson:McClure-Griffiths:2024}.

Fast Radio Bursts provide a uniquely clean measure of baryons via their dispersion measures (DMs), which are equal to the column density of free electrons along the path taken by the burst. When the host galaxy of an FRB is known (the FRB is ``localized''), this can be used to constrain properties of intervening halos and detect invisible baryons in the intergalactic medium. \cite{Macquart:2020} was the first to use the roughly linear relationship between DM and z to detect the so-called ``missing baryons'' \citep{Fukugita:1998}. Analyses by \cite{James:2022b}, \cite{Baptista:2023}, and \cite{Connor:2025} each fitted the distribution of DMs vs redshift in order to estimate the distribution of host galaxy DMs and constrain the strength of feedback in intervening halos. By surveying foreground galaxies and modeling their contributions to the total DM, ``foreground mapping'' techniques aim to achieve tighter model constraints without relying on large samples of FRBs. \cite{Lee:Ata:Khrykin:2022} predicted with simulations that such techniques would be 25$\times$ more constraining than FRB data alone, and that foreground mapping 30 FRB sightlines could constrain the cosmic baryon fraction to better than 10\%. The first data release of the FLIMFLAM survey constrained the fractions of baryons in circumgalactic media (CGM) and IGM to roughly 50\% and 20\% each, as well as estimating the host DM contribution, using only 8 FRB sightlines \citep{Khrykin:Ata:Lee:2024}. \cite{Fujita:Akahori:2017} were among the earliest to propose that FRB DMs could be used to measure the ICM, demonstrating in mock observations that a sufficient number of FRB sightlines combined with SZ measurements could determine the density of the ICM beyond $2 R_{200}$, and be used to derive the temperature profile within $1.5 R_{200}$.

Most foreground mapping studies have utilized a \emph{Modified Navarro-Frenk-White profile} (mNFW) for halo baryon density, introduced by \cite{Prochaska:Zheng:2019} and \cite{Mathews:Prochaska:2017}, to describe all foreground galaxy halos \citep{Simha:2020,Simha:2021,Simha:Lee:2023}. The mNFW profile is typically much flatter at large radii than empirical models of the ICM predict, raising the question of how accurately it can model such a differently shaped halo. \cite{Khrykin:Sorini:2024} demonstrated with SIMBA simulation data that the mNFW model is accurate to within a factor of 10 across a range of halo masses, but stressed the importance of including a mass-dependent baryon fraction. \cite{Huang:2025} adopted an empirically-derived model from \cite{Vikhlinin:2006} to describe halos more massive than $10^{13.5}$~\msun{} while developing their model of the local universe, and \cite{Faber:Ravi:2024} also used this model for an intervening cluster on a single FRB sightline. In contrast, \cite{Connor:2023} worked out a distribution of likely $\DM{ICM}$ values from along two FRB sightlines based on distributions of other components, and compared this to $\DM{ICM}$ from simulated halos, rather than attempt to model the cluster halos themselves.

This paper examines three FRBs from the first CHIME/FRB Outrigger sample \citep{2502.11217} whose host galaxies are members of or lie behind galaxy clusters and groups. One of these is in the infall region of Abell 576 \citep{Rines:2000}, a massive cluster at $z = 0.0382$, far beyond the region of visible X-ray emission, but displaying a surprisingly large DM for its distance. The second FRB host is embedded in Abell 924, a more massive Abell cluster at $z=0.1430$ that shows evidence of substructure near its core \citep{Dupke:Mirabal:2007}. The third FRB host is a member of the group [WH24] J121839.5+484102 (hereafter WH24-J1218) at $z=0.1183$, and lies behind the groups NSCS J121847+484410 (NSCS-J1218) at $z=0.0448$ and [Tully15] Nest 100008 (TullyN08) at $z=0.0024$. All of these are low-mass groups, with no detectable SZ signal, though NSCS-J1218 has a measured X-ray temperature \citep{Lopes:Carvalho:2009}. NSCS-J1218 has been identified as a member of the supercluster MSCC310 by \cite{Santiago-Bautista:2020}, at a junction of cosmic web filaments \citep{Tempel:2014}.

For each of these bursts, we estimate potential contributions to the total extragalactic DM from the intervening cluster or group environments based on three different electron density models. The first is the aforementioned mNFW profile, while the other two are derived from the empirical \emph{Universal Pressure Profile} (UPP) of \cite{Arnaud:2010}, with one assuming an isothermal temperature profile and the other applying an empirical temperature profile fitted to low-Z cool-core galaxy clusters \citep{Baldi:Ettori:2012}. These models were originally fitted to X-ray profiles of the ICMs of low-redshift massive clusters ($10^{14}$ -- $10^{15}$~\msun{}). Subsequently, for FRB 20230703A, we report conservative upper limits of $f_g$ for each model.

This paper is outlined as follows: Section~\ref{sec:frb_sample} provides observational details on the three FRBs under consideration and cataloged measurements of the clusters and groups that their sightlines pass through. Section~\ref{sec:icm_models} describes the ICM/IGrM baryon profile models in detail, and Section~\ref{sec:dm_est} describes the DM calculation from these models. In Section~\ref{sec:results} we estimate the DM contribution to each FRB sightline using these models, using Monte Carlo sampling to account for the uncertainties in cluster masses and concentrations, and constrain the range of $f_g$ allowed by each model. Finally, we discuss our results and make concluding remarks in \ref{sec:discussion}.

We assume Planck 2018 cosmology throughout, with $H_0 = 67.66$~km~s$^{-1}$~Mpc$^{-1}$, $\Omega_m = 0.30966$, and $\Omega_b = 0.04897$ \citep{Planck:2018}, as saved in the \texttt{astropy.cosmology} module \citep{astropy:2013, astropy:2018, astropy:2022}. Cosmology-dependent quantities taken from the literature are converted when necessary.

\section{KKO Sample Cluster Sightlines}
\label{sec:frb_sample}
% Description of the three sightlines intersecting clusters
% Data collection on each cluster
% Mention (after confirming) that these were omitted from the host analysis in Calvin's upcoming paper

% Please add the following required packages to your document preamble:
% \usepackage{booktabs}
% \usepackage{graphicx}

\begin{figure}
	\centering
	\gridline{\fig{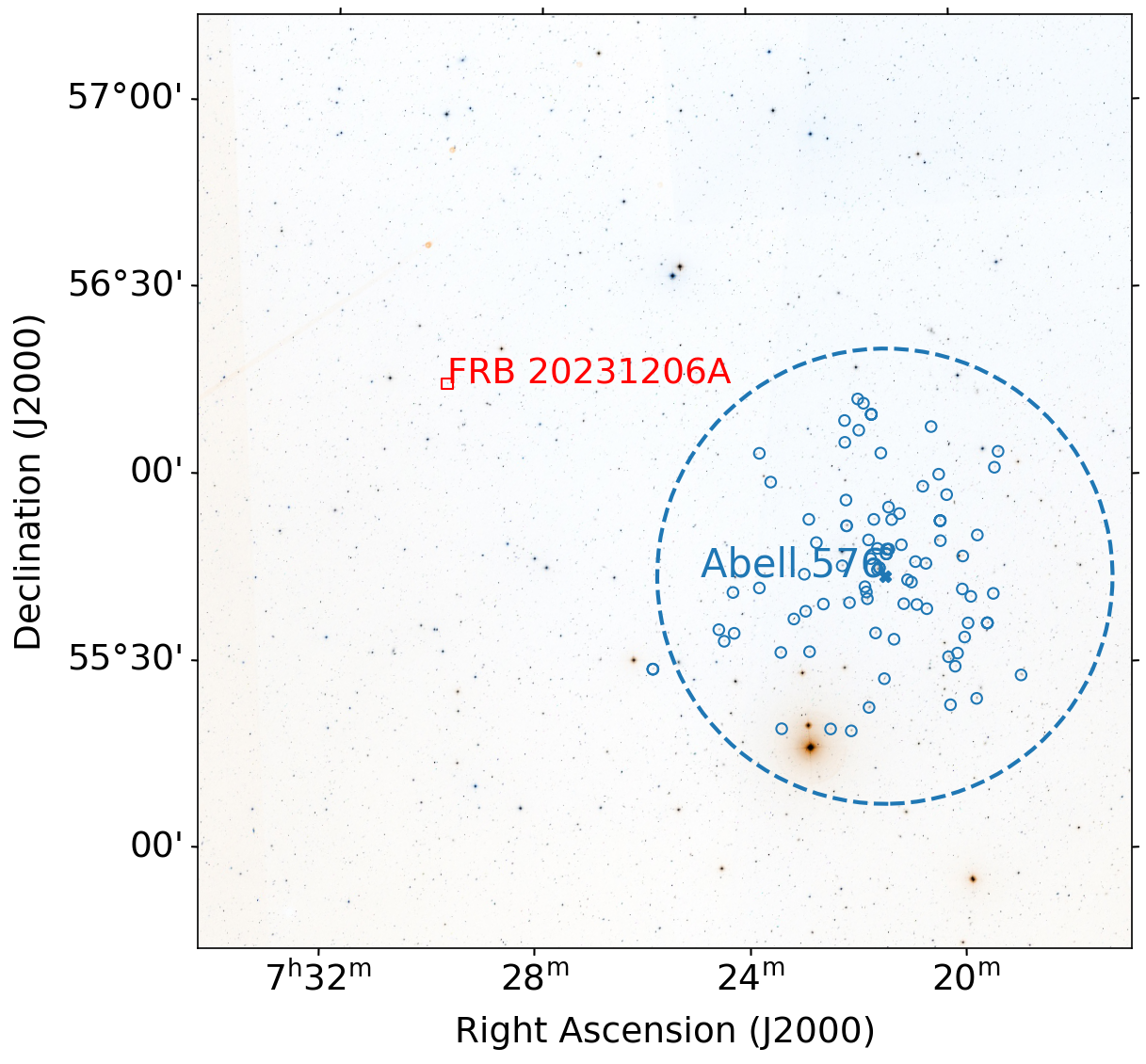}{0.33\textwidth}{(a) FRB 20231206A}
			  \fig{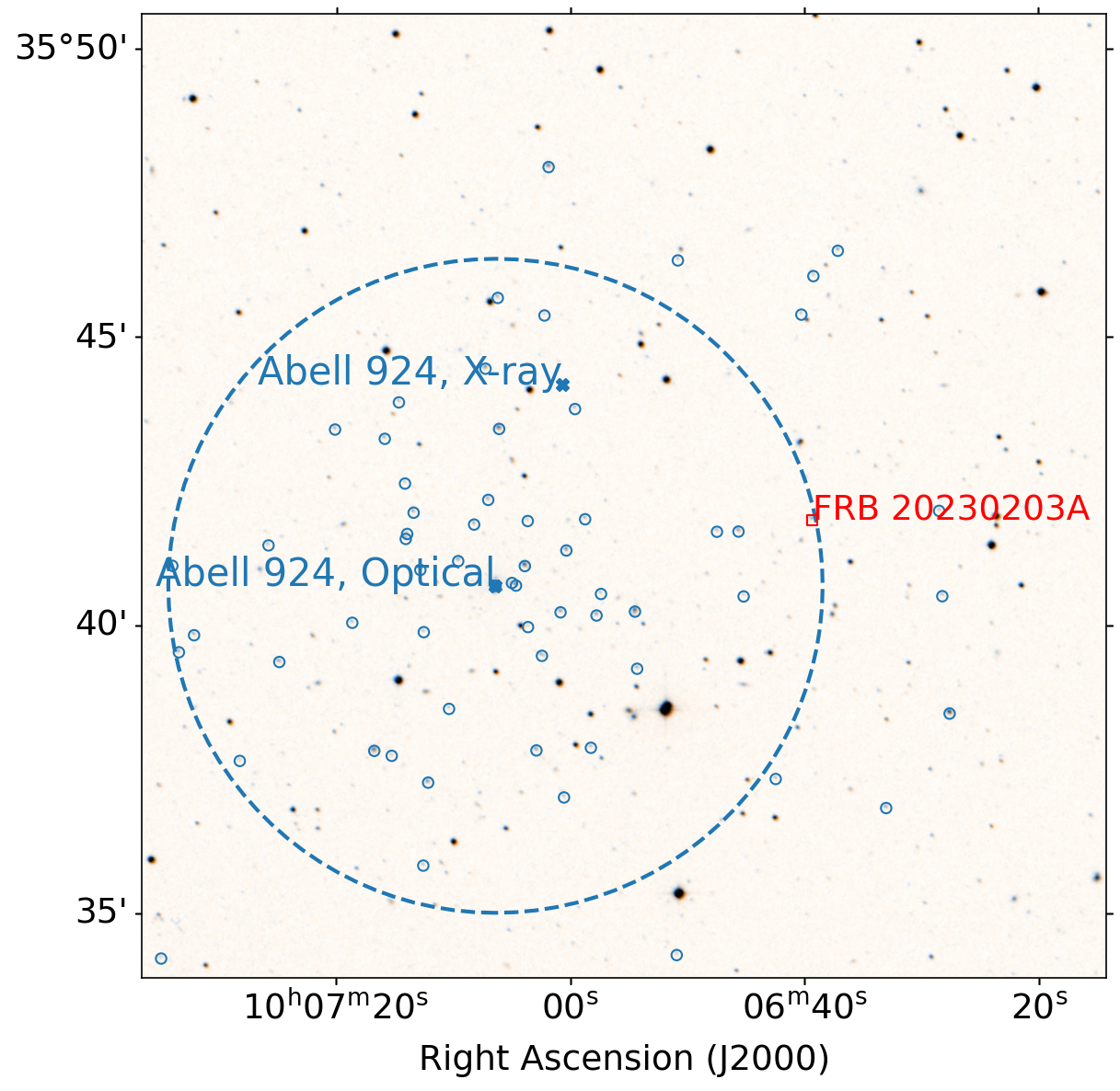}{0.33\textwidth}{(c) FRB 20230203A}
			  \fig{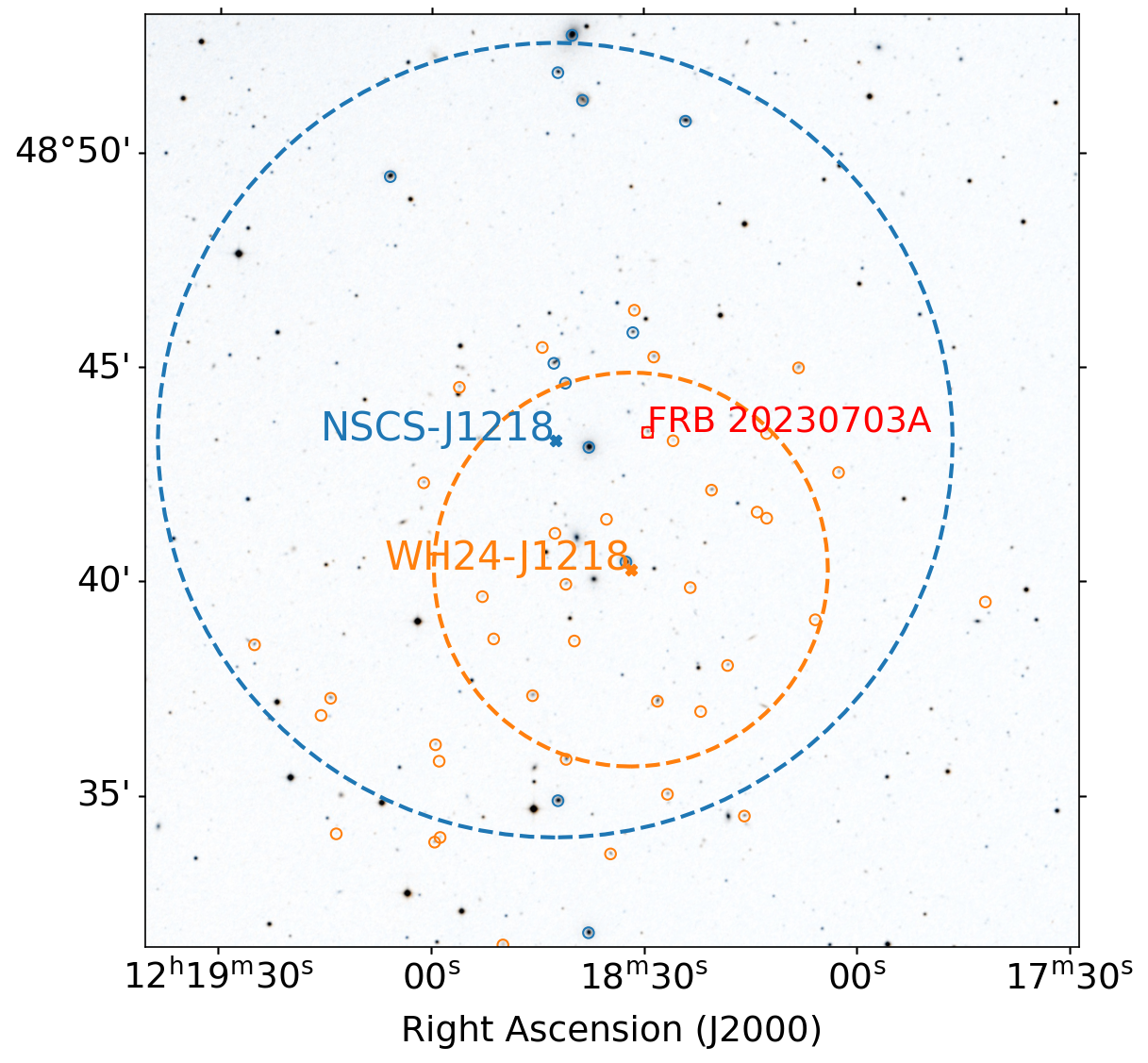}{0.33\textwidth}{(b) FRB 20230703A}
			 }
	\caption{The fields of each FRB host in the Second Digitized Sky Survey (DSS2)\citep{1996ASPC..101...88L}. Member galaxies are indicated by color-coded circles, and the $R_{500}$ virial radii of each cluster is shown by the dashed circles. We omit TullyN08 from the field of FRB 20230703A, since its proximity makes it cover a much larger angle of the sky than the other two groups. For Abell 924, the virial radius is shown around the optical center, but we also plot the X-ray center from \cite{Damsted:2023}. Member galaxies are from Simbad for Abell 576, \cite{Tempel:2012} for NSCS-J1218, \cite{Yang:Xu:He:2021} for WH24-J1218, and \cite{Szabo:2011} for Abell 924}
	\label{fig:fields}
\end{figure}

\begin{table}[]
	\centering
	\hspace*{-1.5cm}
	\resizebox{\columnwidth}{!}{%
		\begin{tabular}{@{}lccc@{}}
			\toprule
			TNS Name        & FRB 20231206A           & FRB 20230703A            & FRB 20230203A           \\ \midrule
			Host name       & 2MASX J07294645+5615122 & SDSS J121829.47+484330.0 & DES J100659.60+354345.3 \\
			Host RA         & 07h29m46.3s             & 12h18m29.5s              & 10h06m39.3s             \\
			Host Dec        & 56d15m11.68s            & 48d43m30.1s              & 35d41m49.7s             \\
			Host z          & 0.0659                  & 0.1184                   & 0.1464                  \\
			\DM{obs} [\dmu{}] & 457.75                  & 290.74                   & 417.3                   \\
			\DM{MW} [\dmu{}]  & 90.46                   & 57.44                    & 65.64                   \\ \bottomrule
		\end{tabular}%
	}
	\caption{FRB host galaxy properties and total DMs in this sample. $\DM{MW}$ is a combination of NE2001 \citep{Cordes:Lazio:2002} for the Milky Way ISM and YT20 \citep{Yamasaki:Totani:2020} for the halo.}
	\label{tab:frbs}
\end{table}

The host galaxies of each FRB in the so-called KKO ``golden sample'' were identified by searching through the Dark Energy Camera Legacy Survey (DECaLS) data release 9 (DR9)  \citep{Dey:Schlegel:Lang:2019} and in the Panoramic Survey Telescope and Rapid Response System (Pan-STARRS) PS1 \citep{Flewelling:Magnier:2019} for non-star objects within the CHIME-KKO localization region, then assessing the likelihood of each object via the Probabilistic Association of Transients with their Hosts (PATH) framework \citep{Aggarwal:2021}. Objects with a PATH probability greater than 90\% were included in the golden sample. We conducted follow-up spectroscopic observations on most of these host galaxies using the Shane spectrograph on the Lick observatory (PI: Kahinga; Prog IDs: 2024A\_S003, 2024B\_S005), the DEIMOS Spectrograph on the Keck 2 Telescope (PI: Prochaska; Prog ID: 2023B-U051), and the Gemini Multi-Object Spectrograph (GMOS) on the Gemini North telescope (PI: Eftekhari; ProgID: GN-2024B-LP-110). Full details of the host associations and optical follow-up are available in the catalog paper \citep{2502.11217}. Table~\ref{tab:frbs} summarizes the properties of the FRBs used in this paper and their host galaxies. The Milky Way DM estimate is a combination of NE2001 \citep{Cordes:Lazio:2002} for the ISM and YT20 \citep{Yamasaki:Totani:2020} for the halo. $\DM{obs}$ refers to the total DM observed by the detector.

In order to model the DM contributions from intervening groups and clusters, we need to identify these halos and find robust measurements of their masses $M_{500}$ and corresponding NFW concentration parameters ($c_{500}$). We began by searching through several large galaxy cluster catalogs to identify candidate foreground halos, first doing a broad cone search of $20^\circ$ and then refining the result to halos that overlap to within a virial radius (as defined in the respective catalog). Subsequent searches in the VizieR database \citep{Ochsenbein:2000} were then done to cross-match candidate groups and clusters across catalogs, and find the best mass estimates. In addition, we searched the SIMBAD service for further information on the host galaxies in the golden sample, which identified two as ``galaxies toward cluster'' (GiC).

The catalogs we searched are:
\begin{itemize}
	\item \cite{Wen:Han:2024} : 1.58 million groups and clusters from DESI Legacy Survey and SDSS-DR10 data. Virial masses were derived from a mass-richness relation ($ M_{500} \geq 4.7 \times 10^{13}$~\msun{}).
	\item \cite{Xu:Ramos-Ceja:2022} : 944 clusters identified in ROSAT All-Sky Survey (RASS) data (RXGCC). Masses were derived from X-ray profiles. ($10^{12}$ -- $10^{15}$~\msun{}).
	\item \cite{Tully:2015} : 25,474 groups and clusters from the Two-Micron All Sky Survey (2MASS), up to $v_\parallel = 10,000$~km/s. Covers 91\% of the sky and a broad mass range (10$^{11}$ -- 10$^{15}$~\msun{}).
	\item \cite{Yang:Xu:He:2021} : 6.4 million groups with more than 3 members, identified from DECaLS-DR9. Virial masses were estimated from total optical luminosity using a halo mass function ($M_\text{halo} \geq 10^{12} h^{-1}$~\msun{}). We keep only groups with more than 10 members from this catalog.
	%	\item \cite{2016A&A...594A..27P} : 1653 SZ-selected clusters from Planck ($10^{13.9} \leq  M_\text{SZ} \leq 10^{15.2}$~\msun{})\footnote{Mass proxy from a scaling relation}
\end{itemize}
In the above list, we also include information on the virial mass estimate used and the approximate mass ranges they cover.

To get the best possible $M_{500}$ for each halo, we gather all available mass estimates for each group and consider the strengths and weaknesses of different mass estimators. For galaxy clusters with resolved X-ray profiles, one can apply conditions of hydrostatic equilibrium to derive both the mass and concentration of the dark matter halo \citep{Vikhlinin:2006,Evrard:Metzler:Navarro:1995}. The self-similarity of clusters leads to several simple scaling relations between observable quantities, such as average X-ray or optical luminosity, or cluster richness, and the virial mass and X-ray temperature \citep{Lovisari:Reiprich:2015}. Several catalogs \citep{Wen:Han:2012,Wen:Han:2024} report masses by calibrating these scaling relations to a handful of resolved X-ray clusters and applying them other groups and clusters. Without X-ray data, the positions and velocities of member galaxies can provide a direct estimate of the halo mass, under the assumption that the member galaxies are in hydrostatic equilibrium. One can solve the Jeans equation to relate the virial mass and concentration to the projected velocity dispersion and galaxy number density as functions of radius, or apply the virial theorem directly to relate the mass to the total projected velocity dispersion \citep[e.g.,][]{Girardi:Giuricin:1998}. The catalogs of \cite{Tully:2015} and \cite{Lopes:Carvalho:2009} report masses derived from the dynamics of member galaxies. For small numbers of galaxies, however, scaling relations have been shown to better recover halo masses in simulations \citep{Marini:Popesso:Dolag:2025}.

Among the sample, the only halo with a published concentration parameter is Abell 576, with $c_{200} = \qty{4.28(0.83:0.81)}{}$ \citep{Babyk:2014}. For the rest of the halos in our sample, we obtain concentration parameteres using M~\textendash~c relations derived from hydrodynamic simulations. Our main results are presented using the cosmology-dependent $M_{500} -- c_{500}$ relation for the Magneticum simulation suite \citep{Ragagnin:Saro:2021}, evaluated at the redshift of each group of cluster. P24 noted that Magneticum was the only simulation among several that did not over-predict the halo gas fraction. Further, the fits in \cite{Ragagnin:Saro:2021} are more flexible to choice of cosmology parameters and virial convention. For comparison, we also ran our analysis using the power-law $M_{200} \textemdash c_{200}$ relations for the IllustrisTNG and EAGLE simulations in \cite{Beltz-Mohrmann:2021}, converting $\Delta = 200$ virial quantities to the $500$ convention by assuming an NFW profile and using the method of \cite{Hu:Kravtsov:2003}.

The following subsections summarize the data obtained from the literature on each sample group or cluster.

\subsection{FRB 20231206A}
% Properties of Abell 576, infall region paper

The SIMBAD listing for ``2MASX J07294645+5615122'', the host galaxy of FRB 20231206A, classifies it as a \emph{galaxy in cluster}, citing \cite{Rines:2000} identifying it as in the \emph{infall region} of Abell 576. Abell 576 is a low-redshift ($z \sim 0.0382$), rich (Abell class 1) cluster, which has been detected optically \citep{Girardi:Giuricin:1998,Abell:1989}, in X-rays \citep{Xu:Ramos-Ceja:2022,Babyk:2014}, and in tSZ \citep{2016A&A...594A..27P}. \cite{Rines:2000} classifies the host as falling toward the core of Abell 576, despite it being well-beyond the limits of its X-ray emitting ICM. At an impact parameter of 3.7~Mpc ($\sim 2.1 R_{500}$) from the X-ray center, we do not expect the sightline of FRB 20231206A to be heavily affected by the ICM of Abell 576.

Dynamical mass estimates for Abell 576 generally skew higher than X-ray derived masses, though estimates for the concentration are varied across methods. The most comprehensive analysis of Abell 576 comes from the Chandra survey of \cite{Babyk:2014}, which resolved the full X-ray profiles of 128 clusters and fitted NFW parameters from them. They derive $M_{200}=\qty{2.111(0.216:0.119)}{} \times 10^{15}$~\msun{} and $c_{200} = \qty{4.28(0.83:0.81)}{}$. We convert these to our $\Delta = 500$ virial convention by taking 20,000 samples of $M_{200}$ assuming masses are log-normally distributed and concentrations are normal, converting each $M_{200}, c_{200}$ sample to $M_{500}, c_{500}$ using the method of \cite{Hu:Kravtsov:2003}, and then finding a log-normal distribution that best fits these values. The result is $M_{500} = \qty{1.474(0.127:0.133)}{} \times 10^{15}$~\msun{} and $c_{500} = \qty{2.80(0.57:0.57)}{}$. In addition, \cite{Babyk:2014} estimate the total baryonic mass of Abell 576 as $M_{b,200}=\qty{2.387(0.214:0211)}{} \times 10^{14}$~\msun{}, which corresponds roughly to $f_{g,200} = \qty{0.72(0.07:0.09)}{}$ (ignoring stellar mass).

These mass estimates are considerably larger than others derived from both dynamical and X-ray means. \cite{Rines:2000} estimate a virial radius of $r_{500} = 1.41\pm0.07$~Mpc, corresponding with $M_{500} = 7.88  \times 10^{14}$~\msun{} (after converting to our fiducial cosmology), and a concentration parameter of $c_{500} = 7.4$. \cite{Wojtak:Lokas:2010} estimate $M_{200}=\qty{6.47(1.04:0.81)}{} \times 10^{14}$~\msun{} and $c_{200} = \qty{3.52(0.61:0.92)}{}$ from the line of sight velocity dispersion, corresponding to $M_{500} = 3.5 $~\textendash~$ 5.2 \times 10^{14}$~\msun{} and $c_{500} = 1.64 $~\textendash~$ 2.69$. \cite{Xu:Ramos-Ceja:2022} estimate, from the X-ray profile, a virial mass of $M_{500} = 1.93\pm0.03 \times 10^{14}$~\msun{}. These discrepancies highlight the challenge of estimating the mass profile of galaxy clusters, even for nearby rich clusters. For our analysis, we use the results of \cite{Babyk:2014}, as these were fitted to resolved X-ray images of the ICM itself.

\subsection{FRB 20230203A}
% Abell 924, subsequently-derived data
Our search through the DESI cluster catalog of \cite{Wen:Han:2024} found the group [WH24] J100706.4+354041 overlapping the line of sight to FRB 20240203A to within $R_{500}$. This cluster was also detected in X-rays as part of the CODEX catalog \citep{Damsted:2023,Finoguenov:2020}, and in multiple SDSS data releases \citep{Rykoff:2016,Wen:Han:2015,Wen:Han:2012,Szabo:2011}, with \cite{Wen:Han:2012} cross-listing it with Abell 924. We will continue to refer to this system as Abell 924, as this is the oldest and shortest identifier.

Member galaxy catalogs are available for this cluster from \cite{Damsted:2023,Rykoff:2016,Szabo:2011}. Among these, only \cite{Szabo:2011} list the host galaxy as a member, likely due to a difference in selection algorithm. \cite{Szabo:2011} used a matched filter technique designed to identify peaks in the galaxy spatial density, while \cite{Damsted:2023} and \cite{Rykoff:2016} used the redMaPPer algorithm \citep{2014ApJ...785..104R}, which identifies clusters as overdensities of red galaxies. Whether or not the host galaxy is a member of the group is only relevant to how we consider the line of sight distance relative to the ICM. The conservative choice is to assume that it is a member galaxy, and consider the range of DMs possible for all line of sight positions.
%  Szabo+11 matched filter tested in [Dong+08 DOI 10.1086/522490]

Notably, \cite{Damsted:2023} report an offset of 3.7' (about 576~kpc) between the X-ray center and the optical center, defined as the position of the brightest cluster galaxy (BCG). Such an offset could be a result of recent merger activity, which would indicate that the cluster may not have fully relaxed yet \citep[e.g.][]{Seppi:2023}. \cite{Damsted:2023} found no evidence of non-Gaussianity in this cluster when applying the Anderson-Darling test of \cite{Hou:Parker:2009}, implying that the galaxies in this cluster are in equilibrium. In contrast, \cite{Wen:Han:2024} characterize this cluster as ``unrelaxed'' according to their $\Gamma$ parameter \citep{Wen:Han:2013}. We do not have a model for a non-equilibrium ICM, so the best we can currently do to account for this is to compute \DM{ICM} relative to both the X-ray center and to the optical center reported by \cite{Damsted:2023}. It turns out that the FRB sightline is at a similar distance from each, and so the difference is negligible compared to other uncertainties. We assume all other physical parameters for this cluster from \cite{Wen:Han:2024}, including a 0.2~dex uncertainty in $M_{500}$.
% , Seppi+23 noticed a trend toward larger offsets in eFEDS clusters which they classified as un-relaxed, consistent with simulations

\subsection{FRB 20230703A}

The host galaxy of FRB 20230703A is behind a lot of structure, presenting an opportunity to constrain total baryon content in a very dense region, but also a challenge in modeling the sightline. As mentioned before, we identify three galaxy groups intersecting the line of sight, and aim to collect the best available measurements for each.

\subsubsection{Host Group}

The group [WH24] J121839.5+484102 (hereafter, ``WH24-J1218'') has a mean redshift and position consistent with the host galaxy of FRB 20230703A \citep{Wen:Han:2024}. This group was previously identified in DESI Legacy Imaging Survey data by \cite{Yang:Xu:He:2021}, who mark the FRB 20230703A host galaxy as a member. This group was also found in SDSS-DR12 \citep{Tempel:Tuvikene:2017}, SDSS-DR8 \citep{Tempel:2012}, and in DPOSS data as part of the Northern Optical Cluster Survey \citep{Lopes:DeCarvalho:2005}.

\cite{Wen:Han:2024} report a virial mass of $M_{500} = \qty{6.8(4.0:2.5)}{} \times 10^{13}$\msun{}, derived from their calibrated mass-richness relation (uncertainties are 0.2~dex). This is roughly consistent with the expectation from applying the $M_{500} $~\textendash~$ L_\text{opt}$ relation of \cite{Popesso:2005} to the total r-band luminosity reported by \cite{Tempel:2012}. We adopt a concentration parameter $c_{500} = 2.64$ from the Magneticum relation \citep{Ragagnin:Saro:2021}. The M--c relations for TNG and EAGLE gives $c_{500}$ values of 2.97 and 3.27, respectively.

\subsubsection{First Foreground Group}

A SIMBAD search along the sightline of FRB 20230703A turned up the galaxy group [AWK2020]-405 of the GalWeight Galaxy Cluster Catalog from SDSS-DR13 \citep{Abdullah:2020}, and subsequent Vizier searches found a corresponding entry in the Northern Sky Optical Cluster Survey \citep{Lopes:Carvalho:2009}, NSCS J121847+484410 (hereafter, ``NSCS-J1218''). This has been identified in a number of optical surveys, but these report a wide range of virial masses from different mass estimation techniques, as listed in Table~\ref{tab:nscs_props} along with corresponding virial radii. The $R_v$ column lists the virial radius under different conventions, while $R_{500}$ gives values corresponding with an overdensity factor of 500. The $R_v$ value from \cite{Santiago-Bautista:2020} corresponds with an overdensity of about 180, which we convert to the 500 convention using a concentration parameter of $c_{180} = 4.47$ from the Magneticum M~\textendash~c relation. The table also shows mass estimates using the \cite{Popesso:2005} scaling relations for total optical luminosity and X-ray temperature, when available.

The measurements from \cite{Abdullah:2020} are a clear outlier from the rest.  Comparing with \cite{Lopes:Carvalho:2009} specifically, it is clear that \cite{Abdullah:2020} selected a larger number of member galaxies with a broader range of line-of-sight velocities. The additional galaxies assigned by \cite{Abdullah:2020} do not appear to be any fainter than those of \cite{Lopes:Carvalho:2009}, so the difference is likely due to the group-membership algorithm used. \cite{Abdullah:2020} assign galaxy membership using the sophisticated \emph{GalWeight} scheme, which takes into account galaxy density in the position/velocity phase space as well as distributions in their projected positions and line of sight velocities \citep{Abdullah:Wilson:2018}, while \cite{Lopes:Carvalho:2009} identified groups in redshift space alone, using a velocity gap technique. Although GalWeight is well-validated against simulations, especially for isolated group, it is possible that it may be too permissive for small groups embedded in larger systems. Given the preponderance of literature favoring a lower-mass for this system, we choose to use the measurements of \cite{Lopes:Carvalho:2009} for our DM estimates. The error bars there are reported as 68\% confidence limits, and we assume the masses are log-normally distributed.

\begin{figure}[h]
	\centering
	\includegraphics[width=1.0\linewidth]{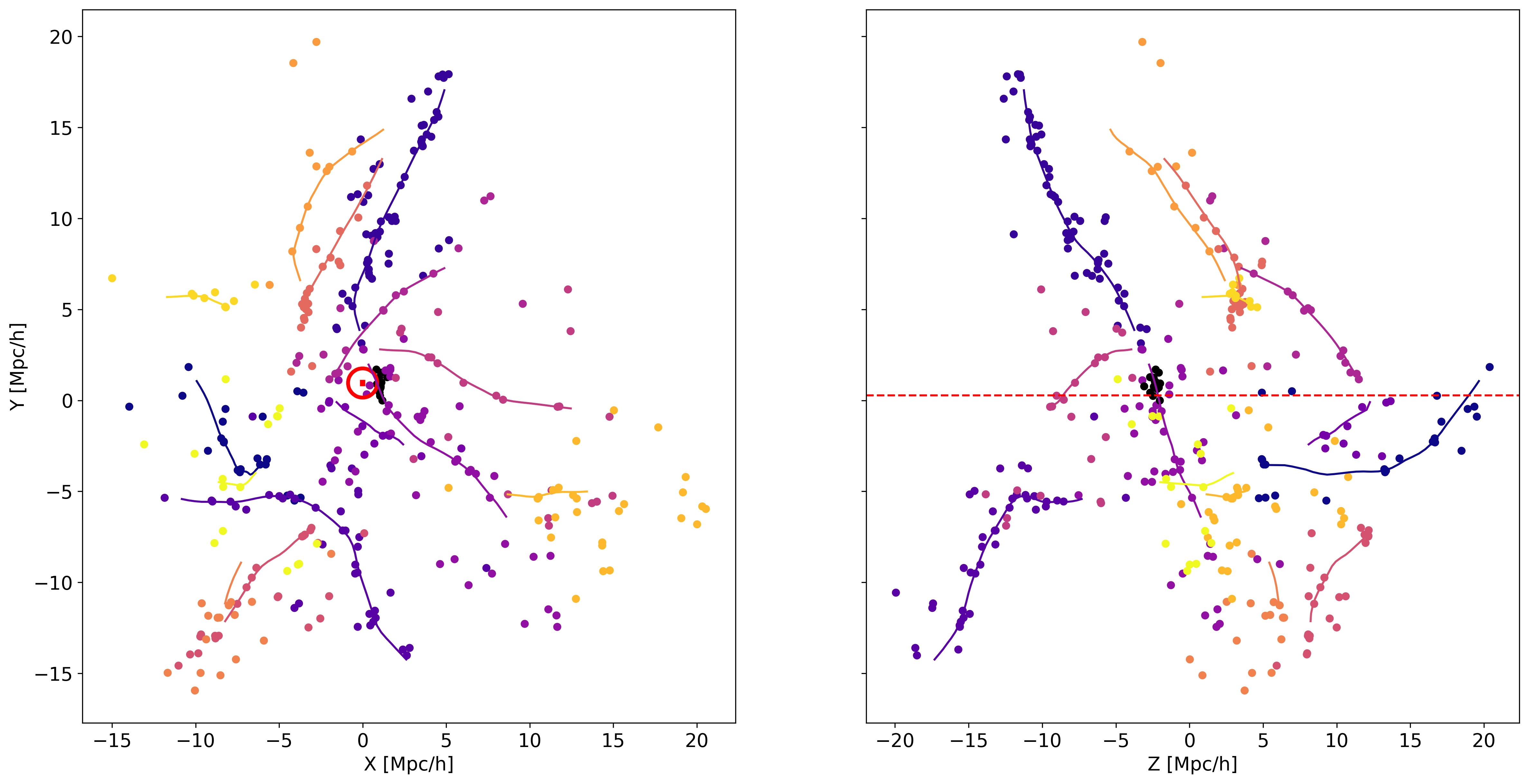}
	\caption{Comoving positions of galaxies (points) and filaments (lines) within 10~Mpc/h of the center of NSCS-J1218, from \cite{Tempel:2014}. Galaxies are color-coded by the nearest filament. Galaxies that are members of group 15955 of \cite{Tempel:2012}, which is equivalent to NSCS-J1218, are marked in black. Cartesian positions are relative to the group center, with the Z direction along the line of sight to FRB 20230703A, and X and Y orthogonal. The left panel shows the XY plane, and the right shows the ZY plane. The red dotted circle on the left and the red dashed line on the right show the line of sight.}
	\label{fig:0703_slice}
\end{figure}

\cite{Santiago-Bautista:2020} identified NSCS-J1218 as part of the supercluster MSCC-310. Likewise, \cite{Tempel:2014} found this group to be at a juncture of several apparent cosmic filaments, as shown in Fig.~\ref{fig:0703_slice}. This may imply that the group is actively accreting material from its surroundings, and may not be fully relaxed. Other observations of this group have indicated some degree of anisotropy -- member galaxies from \cite{Tempel:2012} (marked in black in Fig.~\ref{fig:0703_slice}) extend about twice as far in Dec as in RA. \cite{Lopes:Carvalho:2009} find marginally significant anisotropy in NSCS-J1218 with their $\beta$ test. Since our sightline passes very near the centroid position of the group (and, presumably, the intragroup medium), we will assume that a spherical halo is a good approximation.

We note that NSCS-J1218 was not detected by \cite{Wen:Han:2024}, despite having a richness above their detection cutoff ($N_\text{gal}(<r_{500}) \geq 6$). The most likely reason for this is that NSCS-J1218 does not appear to have a BCG-like member, which was the basis for group discovery in the \cite{Wen:Han:2024} catalog.

\begin{table}
	\hspace*{-0.7in}
\begin{tabular}{lcccccc}\toprule
	& $N_\text{gal}$ & $R_v$ & $R_{500}$ & $M_{500}$ & $M_{500,\text{opt}}$ & $M_{500,\text{x-ray}}$\\
	&  & [Mpc/h] & [Mpc/h] & \multicolumn{3}{c}{[$10^{14}$ $M_\odot/h$]} \\
	\hline \hline
	\cite{Lopes:Carvalho:2009}  &  $12 \pm 3$  &  --  &  $\qty{0.34(0.08:0.04)}{}$  &   $\qty{0.24(0.16:0.10)}{}$  & $\qty{0.15(0.07:0.05)}{}$  &  0.16 \\
	\cite{Tempel:2012}  &  17  &  0.25  &  --  &  --  &  $\qty{0.21(0.10:0.07)}{}$  &  -- \\
	\cite{Santiago-Bautista:2020}  &  15  &  0.35  &  0.22*  &  0.06*  &  --  &  -- \\
	\cite{Abdullah:2020}  &  21  &  --  &  0.54  &  $1.00 \pm 0.37$  &  --  &  -- \\
	\bottomrule
\end{tabular}
\caption{Properties of NSCS-J1218, from different sources. $N_\text{gal}$ is the number of galaxies identified by the group finder, $R_v$ is a virial radius calculated from velocity dispersion, and $R_{500}$ is the standard virial radius when given. When total optical luminosity or X-ray temperature are available from a given source, the result of the \cite{Popesso:2005} scaling relations are shown in the last two columns. Starred quantities were converted from the reported $R_v$ value assuming an NFW profile with concentration parameter $c_v = 4.47$ (derived from the Magneticum simulations, \cite{Ragagnin:Saro:2021}).}
\label{tab:nscs_props}
\end{table}

\subsubsection{Second Foreground Group}

A search through the \cite{Tully:2015} catalog found the group [T2015] Nest 100008 (herafter ``TullyN08'') in the foreground, at an impact parameter of 719~kpc and a distance of 15~Mpc ($z \sim 0.003$). \cite{Tully:2015} quantifies group radii in terms of the projected radius of second-turnaround ($R_{2t}$), the maximum radius at which galaxies that have already fallen through a group core will reach as they orbit outward. This quantity is slightly larger than the theoretical virial radius, and appears more naturally as a cusp in velocity space. \cite{Tully:2015} identified groups from the 2MASS redshift survey by iteratively collecting galaxies within $R_{2t}$ of a candidate group, updating the total luminosity, and re-computing $R_{2t}$, at each step using empirical mass -- light and $R_{2t}$ -- mass relations.

In addition to the luminosity-mass estimate, Tully calculates a virial mass dynamically using the formulae in \cite{Tully:2015a}, which for TullyN08 is $M_v = 5.5 \times 10^{13} M_\odot / h$, corresponding with a line-of-sight velocity dispersion $\sigma_P = 366$~km~s$^{-1}$, estimated from $N_\text{gal} = 65$ member galaxies. To convert these parameters to the $M_{500}$ convention and our assumed cosmology, we apply the following equation from \cite{Lopes:Carvalho:2009}:
\begin{equation}
	R_\Delta = R_A (\rho_V / (\Delta \rho_c))^{1/2.4},
\end{equation}
where we take the aperture radius $R_A = R_{2t} = 745$~kpc, $\Delta = 500$, $\rho_V = 3 M_v / (4 \pi R_{2t}^3)$, and $\rho_c$ is the critical density. The exponent is approximately accurate at the virial radius for an NFW profile. To estimate the uncertainty in the mass, we bootstrap the member galaxy data published with the catalog, recomputing $M_v$ 5000 times with random resampling and converting each to $M_{500}$. The result is $M_{500} = \qty{6.8(2.3:1.9)}{} \times 10^{13}$~\msun{}. The bootstrapped $M_{500}$ values are described well by a $\Gamma$ distribution (Kolmogorov-Smirnov test statistic value of 0.0065) so we sample from this distribution in our Monte Carlo analysis.

% Conclusion from comparing Abdullah+20 and Lopes+09 member galaxies:
%	 Abdullah+20 includes a wider range of LOS velocities, hence gets a larger mass. No noticeable difference in member galaxy fluxes, so it's not a matter of limiting magnitude. Lopes+09 utilized shifting gapper technique to identify group members. Both compute mass from radial velocity. Lopes+09 give Lopt = 2.29e11 Lsol, which corresponds with M200 = 4.16e13 msun{} by the Popesso+07 Mass/Lum relation. Don't have a total optical luminosity from the Abdullah sample. Estimating from published r_band magnitudes, I get L_r = 4.5e11 --> 8.6e13 msun{}

\section{ICM / IGrM Models}
\label{sec:icm_models}
%To estimate the contribution of a foreground halo to an FRB's dispersion, we need a model of the electron density within the halo. 
While the total mass density is commonly described by the Navarro, Frenk, and White profile (NFW) \citep{Navarro:Frenk:White:1997}, the baryons (and hence, free electrons) take on a different structure due to hydrodynamic forces. The mNFW profile is commonly used for galaxy-scale halos, as it is a flexible model intended to describe the feedback-driven migration of material from the inner to outer regions of the halo. ICM models are derived from conditions of hydrostatic equilibrium, and are usually written to relate observables like X-ray luminosity and galaxy velocity dispersion to the thermodynamic properties of the gas. Due to the self-similarity of massive clusters, density profiles that depend only on the total mass can fit fairly well to observations, but are typically not extrapolated beyond $R_{500}$. In this section, we will discuss the Modified NFW in more detail, as well as the empirical UPP for massive cluster ICMs.

%Each model presented here is proportional to a parameter $f_\text{hot}$, describing the fraction of baryons in the halo that are not collapsed into stars or un-ionized. $f_\text{hot}$ is given relative to the global baryon fraction $f_B$, such that
%\begin{equation}
%	f_\text{hot} = f_g - f_* - f_\text{cold},
%\end{equation}
%where $f_g$ is the total baryon fraction per Eq.~\ref{eq:fg_def}, $f_*$ is the mass fraction collapsed into stars, and $f_\text{cold}$ is the fraction in cold gas. For group and cluster mass halos, $f_*$ and $f_\text{cold}$ are very small. When calculating DM contributions, we will assume $f_\text{hot} = f_g = 1$ for all halos, and then use over-predictions of DM to set upper limits on $f_g$.

\begin{figure}
	\centering
	\includegraphics[width=1.0\linewidth]{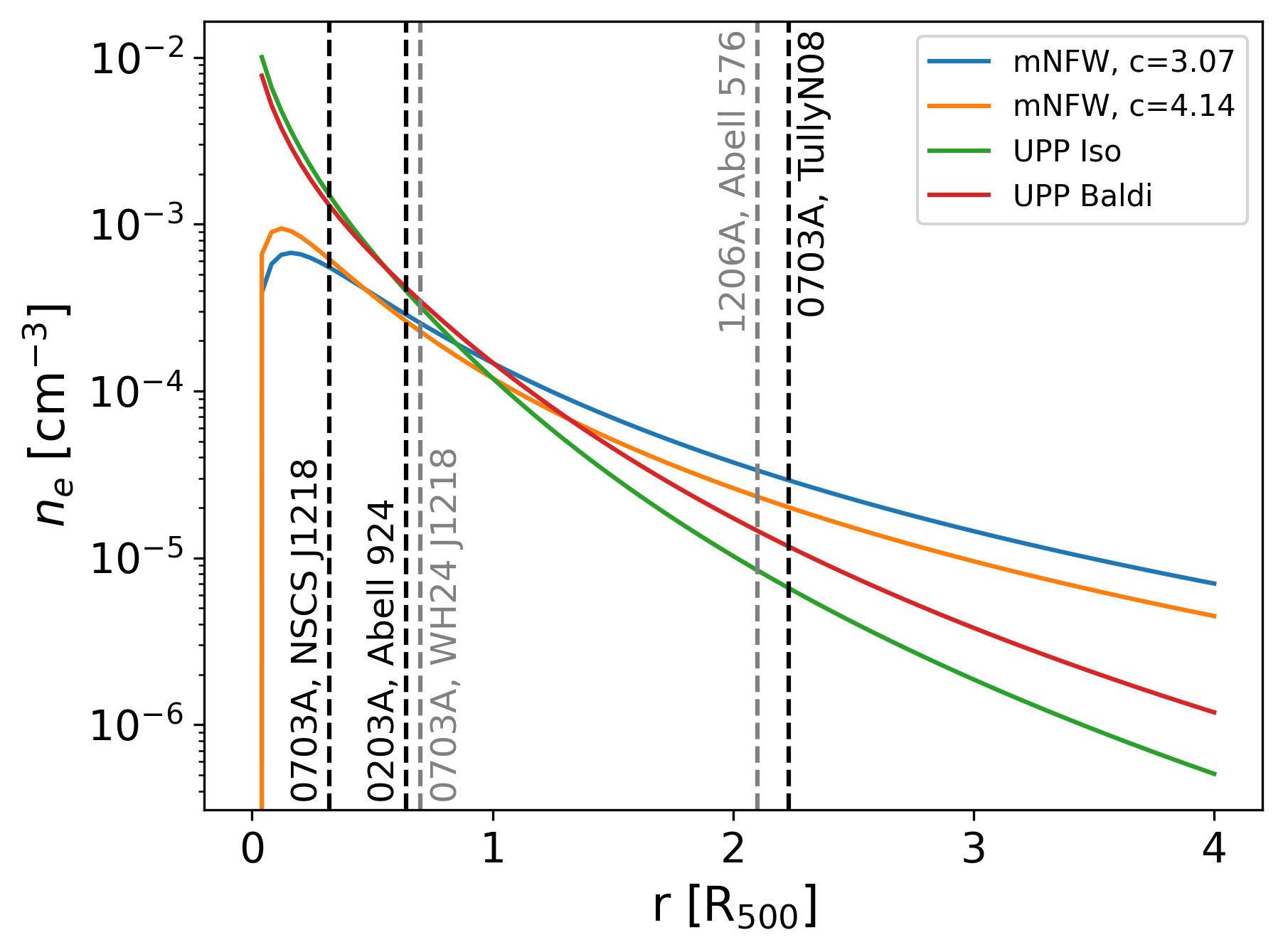}
	\caption{Comparison of ICM electron density models for $f_{g,500} = 1$ and $M_{500} = 10^{15}$~\msun{}. Curves for the mNFW profile are shown for different concentration parameters. The two values (3.07 and 4.14) correspond with halo masses of $10^{14}$ and $10^{15}$~\msun{} using the Magneticum $M_v $~\textendash~$ c$, respectively, though the curve for each is still shown for $10^{14}$~\msun{}. Vertical lines mark the impact parameters of each FRB (indicated by the last five characters of its TNS name) with each group or cluster in the sample.}
	\label{fig:compare}
\end{figure}

\subsection{Modified NFW}

The mNFW baryon density profile is defined as
\begin{equation}
	\rho_b(y) = \frac{f_g(R_v) f_B \rho_0}{(c y)^{1-\alpha}(y_0 + c y)^{2+\alpha} },
	\label{eq:mnfw}
\end{equation}
where $y = r / R_v$ is the radial position normalized to the virial radius $R_v$, and $c$ is the NFW concentration parameter. The density $\rho_0 = M g_{b}(c) / (4 \pi (R_v/c)^3)$, where the function $g_b(y)$ characterizes the fraction of mass contained within a given radius. The electron number density is $n_e(y) = \rho_b(y) \times \mu_e / (\mu m_p)$, where $\mu_e = 1.1667$ is the number of free electrons per hydrogen assuming full ionization a 25\% helium mass fraction, and $\mu = 1.33$ is the mean molecular weight per hydrogen. For FRB foreground modeling, $y_0 = \alpha = 2$ is typically assumed, as this closesly matches earlier models of galactic halos \citep{Prochaska:Zheng:2019}. The resulting profile increases from the core out to twice the NFW scale radius, and then drops off as $r^{-3}$ at large radii. 

In Eq.~\ref{eq:mnfw}, we've included $f_g$ as a parameter which depends on the virial radius convention used. This is equivalent to the $f_\text{hot}$ parameter in other literature and in the \texttt{frb} Python package\footnote{\url{http://github.com/FRBs/frb}}, except here we've made its dependence on halo radius explicit. In the \texttt{frb} Python package, the default convention uses the formula from \cite{Bryan:Norman:1998} (denoted by subscript ``v'' here, see App~\ref{app:virial_conv}). We choose to use the $\Delta=500$ convention here for consistency with the UPP model (see next section). Under the default virial convention, $f_\text{hot}$ is usually chosen to be about $0.9$ for cluster-mass halos and $0.75$ for galaxy-scale halos \citep[e.g.][]{Lee:Khrykin:2023}.

Despite its success for modeling CGMs, mNFW's ability to model ICM gas has yet to be demonstrated. \cite{Lee:Khrykin:2023} used it to model the contributions of two clusters in the foreground of FRB 20190520B, and found it consistent with the total observed DM. \cite{Huang:2025} applied the modified NFW model only for halos with masses less than $10^{13.5}$~\msun{}, and used a scaled version of the model fitted to Abell 907 by \cite{Vikhlinin:2006} for more massive halos. The \cite{Vikhlinin:2006} model has a very flexible form of multiple power laws, which they fitted to Chandra observations of massive clusters. While it provides a good fit for each cluster profile, it is not guaranteed that the fit to a given cluster in their sample should scale naturally to others.

\subsection{Universal Pressure Profile}
For comparison we derive two electron density models from the UPP \citep{Arnaud:2010}, which has well-motivated scaling with mass and redshift. Among averaged scaled quantities, the pressure is the least affected by dynamical history and feedback effects \citep{Arnaud:2010}. The UPP was derived by fitting X-ray observations of the REXCESS sample of galaxy clusters ($z < 0.2$, $10^{14} < M_{500} / M_\odot < 10^{15}$) to a generalized NFW form, after normalizing to characteristic values based on the virial mass of each. The result is
\begin{align}
	p(y) &= \left( \frac{M_{500}}{4.28 \times 10^{14} h^{-1} M_\odot} \right)^{\alpha_P(y)}  \frac{P_0}{(c_{500} y)^{\gamma} (1 + (c_{500} y)^\alpha)^{(\beta - \gamma)/\alpha}} \label{eq:upp_general} \\
	\alpha_P(y) &= 0.22 \left(1 - \frac{(y/0.5)^3}{1 + (y/0.5)^3}\right) \\
	P(y) &= P_{500} p(y).
\end{align}
$P_{500} = n_{e,500} k_B T_{500}$ is the characteristic pressure of the halo, $n_{e,500} = 500 \rho_c(z) / (\mu_e m_p)$ is a characteristic electron density, and $\rho_c$ is the critical density. The exponent function $\alpha_P$ comes from the relationship between an SZ-effect mass proxy and $M_{500}$ used by \cite{Arnaud:2010}. The characteristic temperature is taken as
\begin{equation}
	k_B T_{500} = \mu m_p G M_{500} / 2 R_{500},
	\label{eq:t500_iso}
\end{equation}
which is the temperature of a singular isothermal sphere, with $\mu = 0.6$ the mean molecular weight of primordial gas. The best-fit parameters are $[P_0, c_{500}, \gamma, \alpha, \beta] = [4.980 h^{-3/2}, 1.177, 0.3081, 1.0510, 5.4905]$. The result is a profile that drops off as $\sim r^{-5}$ at large radii, much faster than the mNFW profile.

We note that \cite{2015MNRAS.451.3868L} compared simulated cluster ICMs across a wide range of mass bins (down to $10^{12}$~\msun{}) and found that pressure profiles drop off less steeply at large radii (r > $r_{500}$) than the UPP predicts. They also found that the UPP overestimates the pressure at small radii for low mass-halos with feedback. \cite{2015MNRAS.451.3868L} point out that their simulated pressure profiles are in agreement with tSZ pressure profile meausurements from Planck out to large radii \citep{2013A&A...550A.131P}.

To convert the pressure profile to a density profile, we apply two different temperature profiles. The simplest is to assume the halo gas is isothermal with temperature $T_{500}$ given in Eq~\ref{eq:t500_iso}, which is a decent approximation to start. For a more realistic temperature model, we use the power-law profile from \cite{Baldi:Ettori:2012} fitted to the outer ($r > 0.15R_{500}$) regions of XMM-Newton clusters:
\begin{equation}
	\frac{T(y)}{T_{500}} = \frac{A}{(1 + (y/0.4)^2)^\beta_0}.
	\label{eq:baldi}
\end{equation}
The authors fitted this form to data in three redshift bins, with the lowest bin ($0.1 < z < 0.2$) using data for 44 clusters from \cite{Leccardi:Molendi:2008}, splitting the sample into cool-core (CC) and non-cool-core (NCC). The difference between the two is less than 16\% up to $4 R_{500}$, so we assume the CC model here. Table~8  of \cite{Baldi:Ettori:2012} gives $A = 1.09$ and $\beta_0 = 0.24$.

From the UPP and each temperature profile, we obtain the electron number density as
\begin{equation}
	n_e(y) = P(y) / k_B T(y), \label{eq:ne_upp_raw}
\end{equation}
where $T(y)$ is given by the CC fit from \cite{Baldi:Ettori:2012} for the ``UPP Baldi'' model, and $T(y) = T_{500}$ for the ``UPP Iso'' model.	

As empirical fits, the UPP-derived models implicitly assume a gas fraction, which we would like to be a tunable parameter. To correct this, we integrate Eq.~\ref{eq:ne_upp_raw} numerically to calculate the implicit gas fraction $\hat{n}_0$ and then normalize the expression by this (see Appendix~\ref{app:upp_norm}). The result is
\begin{equation}
	n_e(y) = f_g(R_v) \left(\frac{f_B \rho_{500} T_{500}}{\hat{n}_0 \mu_e m_p}\right) \frac{p(y)}{T(y)}.
	\label{eq:ne_upp}
\end{equation}
Mass dependence is included via the power-law factor in $p(y)$ and in the definition of $T_{500}$. Redshift-dependence comes from $\rho_{500}$, which depends on the critical density.

\section{$\DM{ICM}$}
\label{sec:dm_est}

\begin{figure}
	\centering
	\includegraphics[width=0.5\linewidth]{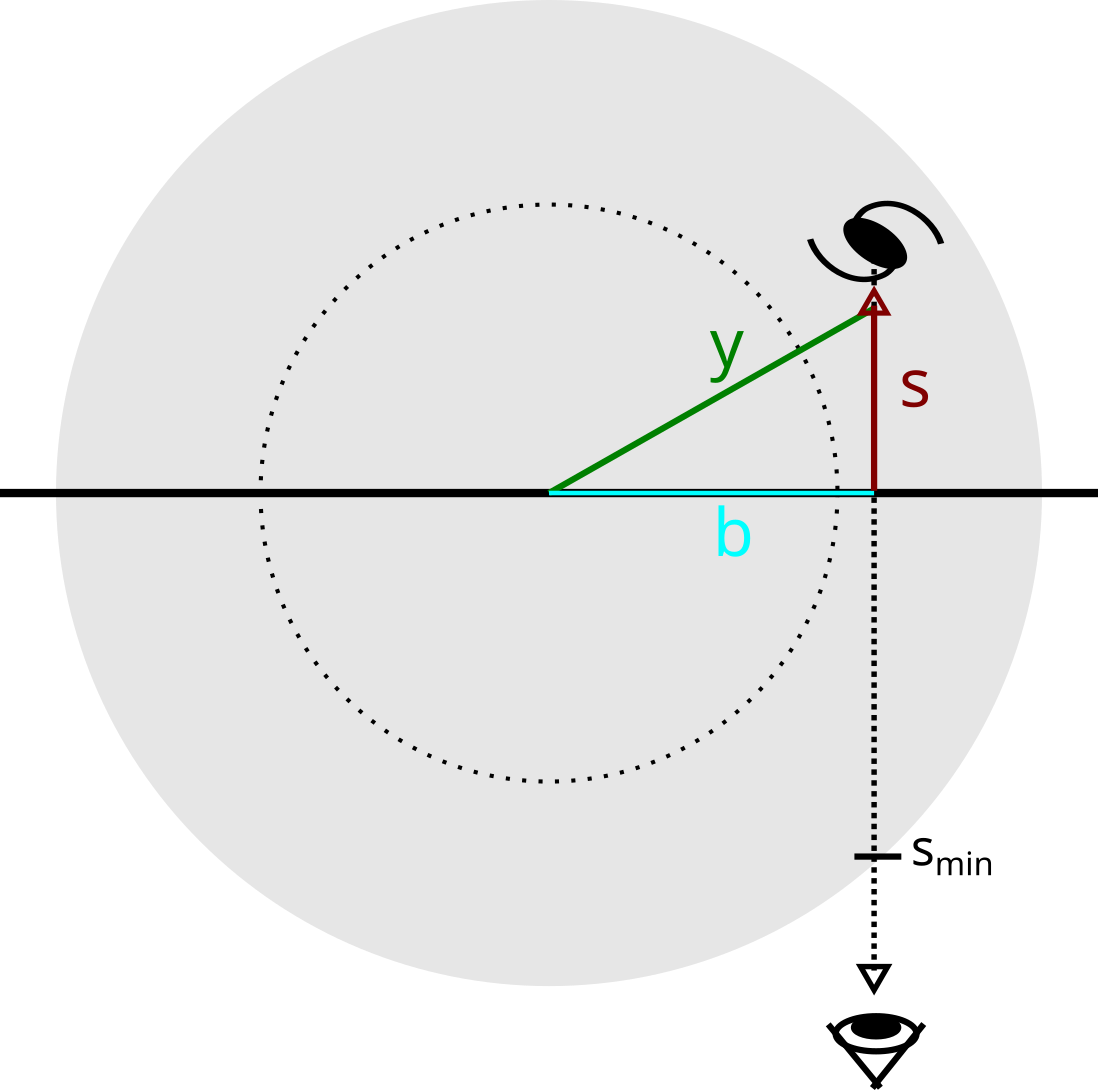}
	\caption{Halocentric coordinates to for computing $\DM{ICM}$. The impact factor ($b$), radius ($y$), and line of sight position parameter ($s$) are all kept in units of $R_{500}$. The integral is taken from the nearest point on the sightline that is within the halo (denoted by $s_\text{min}$), to the each sampled host galaxy position. When the host galaxy is confidently behind the halo, the integral is taken to $s_\text{max} = -s_\text{min}$.}
	\label{fig:coorddiag}
\end{figure}

The DM contribution of each halo is computed numerically as
\begin{equation}
	\DM{ICM} = \int\limits_{s_\text{min}}^{s_\text{max}} n_e(\sqrt{s^2 + b^2}) ds,
	\label{eq:dm_icm_def}
\end{equation}
where $s$ is the line-of-sight position within the halo, relative to the point of closest approach, and $b$ is the impact parameter (see Fig.~\ref{fig:coorddiag}). The integral is taken from $s_\text{min} = -\sqrt{y_\text{max}^2 - b^2}$, where $y_\text{max}$ is a chosen maximum radius in units of $R_{500}$.

For each model and halo, we generate 10,000 halo realizations, sampling over uncertainties in mass, concentration, and (if the host is a member of the cluster) depth $s_\text{max}$ (see Sec.~\ref{sec:depth_dist}). For TullyN08, we assume masses follow a gamma distribution, since that was the distribution most consistent with bootstrapped mass estimates. For all others, we assume lognormally-distributed values. To account for halo-to-halo uncertainty in $c_{500}$, we convert the fiducial value of $c_{500}$ to $c_{200}$ and draw lognormal samples with a scatter of 0.22~dex, which is a fairly conservative assumption based on the OmegaWINGS survey of \cite{Biviano:Moretti:2017}. The samples are then converted back to $c_{500}$. Uncertainty in $c_{500}$ due to mass variation is only around 2\% of the uncertainty due to halo-to-halo variation.

\begin{figure}
	\centering
	\includegraphics[width=0.7\linewidth]{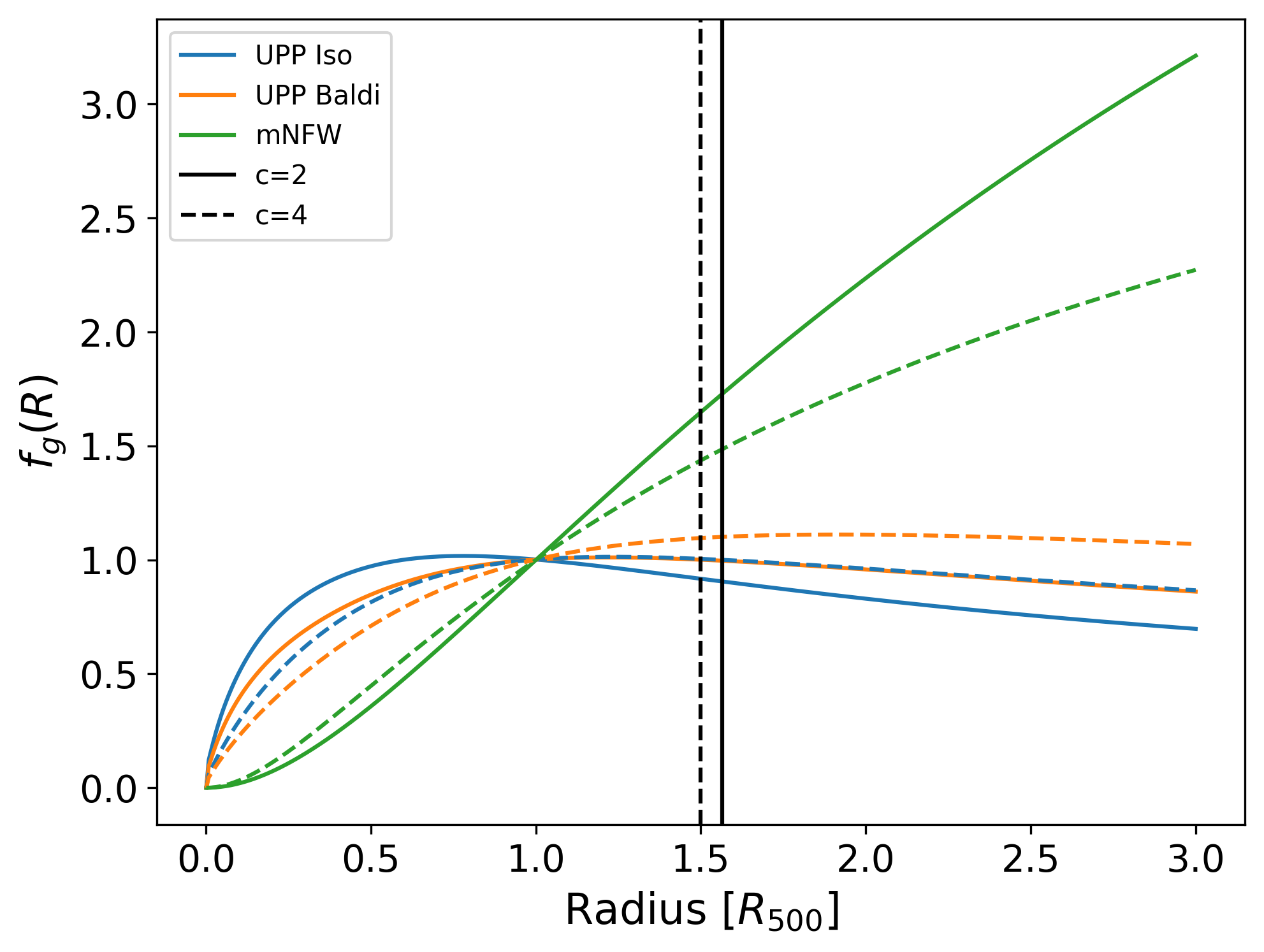}
	\caption{Gas fractions vs. radius for each model for different halo concentrations, normalized so that $f_{g,500} = 1$. The vertical black lines mark the locations of $R_{200}$, which depends on the concentration parameter. We use the $c_{500}$ and $z$ values listed in Table~\ref{tab:clusts} to correct DMs integrated beyond $R_{500}$ to different virial radius definitions.}
	\label{fig:mass_ratios}
\end{figure}

We would like to use the \DM{ICM} estimates from Eq.~\ref{eq:dm_icm_def} and the total DMs for the FRBs in our sample to constrain $f_{g,500}$ and $f_{g,200}$. However, each of the models in Sec.~\ref{sec:icm_models} is normalized such that $f_{g,500} = 1$. If we take the integral beyond $R_{500}$, the constraints from each model are no longer comparable, since the value of $R_{200}$ depends on the halo's concentration and redshift. Fig~\ref{fig:mass_ratios} shows the value of $f_g(r)$ for each model for the same halo mass at two different concentrations, with the position of $R_{200}$ marked by black lines. To re-normalize so that $f_{g}(R) = 1$ for some maximum radius $R$, we scale the DM estimate by $f_B M_v/M_b(R)$, where $M_b(R)$ is the baryonic mass contained within $R$.

\subsection{Depth Sampling}
\label{sec:depth_dist}

For an FRB host galaxy that is a member of a cluster or group, how much the ICM/IGrM contributes to the DM will depend on its depth within the cluster. The galaxy redshift $z_g$ does not provide this distance information, since peculiar motion within the cluster dominates over the Hubble flow on this scale. If we assume that the cluster mass distribution follows an NFW profile and that galaxy positions trace dark matter, then the probability density function for galaxy position along the line of sight is proportional to the NFW profile, restricted to the chord through the spherical halo:
\begin{equation}
	\text{PDF}(s) = \frac{A}{(c y(s)) (1 + c y(s))^2} \qquad y=\sqrt{b^2 + s^2}
\end{equation}
where $A$ is a normalization constant such that $\int \text{PDF}(s) ds = 1$.

\section{Results}
\label{sec:results}

\begin{figure}[h]
	\centering
	\includegraphics[width=1.0\linewidth]{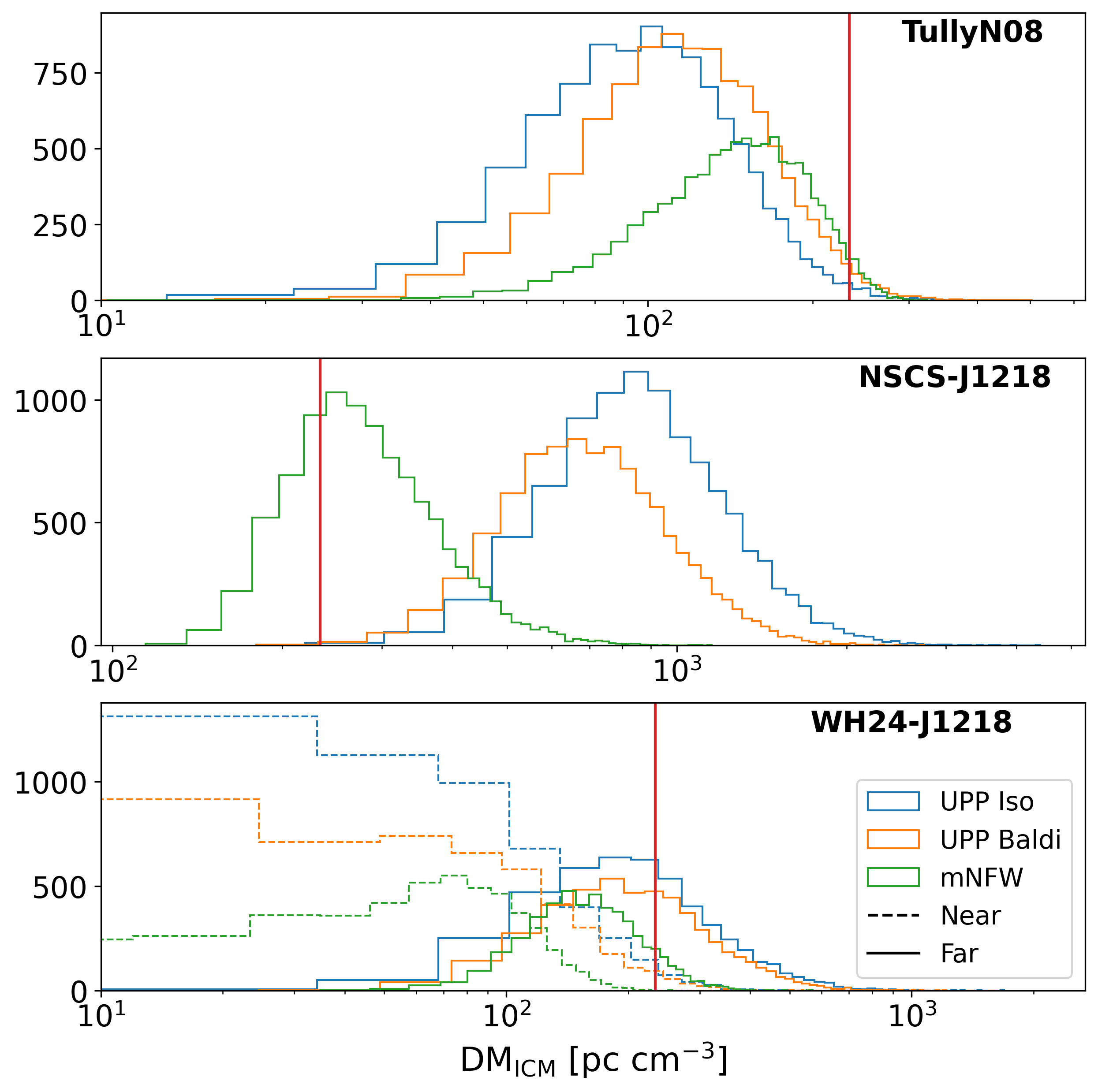}
	\caption{Distributions of \DM{ICM} for each model for each group along the sightline of FRB 20230703A. For WH24-J1218, samples are taken over line-of-sight distance $s_\text{max}$ within the group. DMs for distances  $s_\text{max}>0$ (on the ``far side'' of the cluster center) are histogrammed with a solid line, while ``near side'' samples ($s_\text{max}< 0$) are plotted with a dashed line.  The x-axis is set to a lower-bound of 10~\dmu{} for a cleaner display, but around 2 \textendash 5\% of DM samples fall below this threshold. The vertical red lines show the total extragalactic DM of FRB 20230703A.}
	\label{fig:burst1_dmicm}
\end{figure}

\begin{figure}
	\centering
	\includegraphics[width=1.0\linewidth]{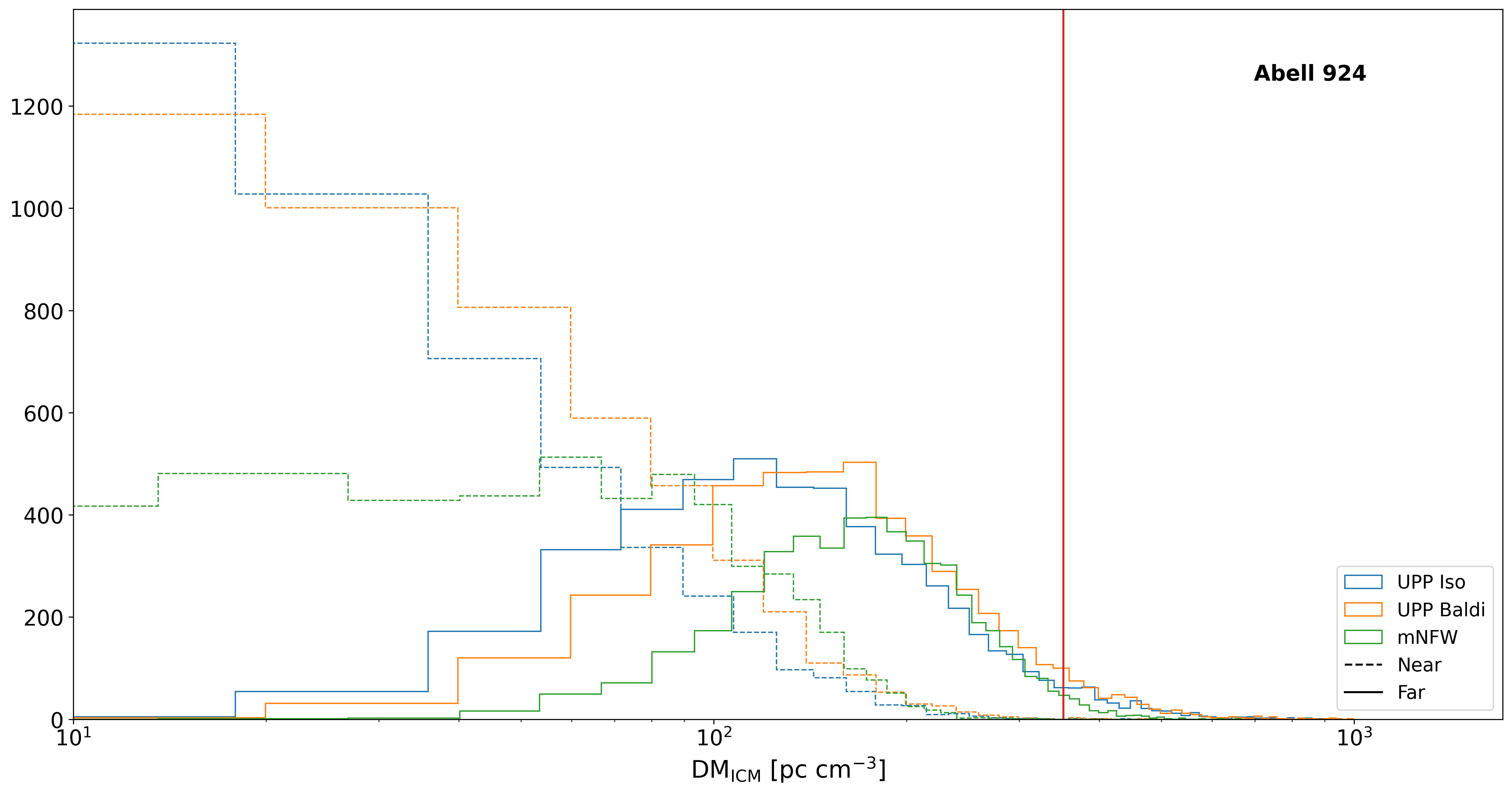}
	\caption{Distributions of \DM{ICM} for each model for the sightline of FRB 20230203A. As in Fig.~\ref{fig:burst1_dmicm}, near-side line of sight samples are histogrammed with a dashed line, and far-side samples with a solid line. The difference between the two is negligible compared to uncertainties from line-of-sight position, halo mass, and concentration. The vertical red line shows the total extragalactic DM of FRB 20230203A. Around 7\% of samples fall below the 10~\dmu{} cutoff on the x-axis}
	\label{fig:burst2_dmicm}
\end{figure}

Fig.~\ref{fig:burst1_dmicm} shows histograms of the DM estimates for halos along the FRB 20230703A sightline, scaled by redshift to the observer's frame.  Fig.~\ref{fig:burst2_dmicm} shows similar histograms for the FRB 20230203A sightline using the optical center. The median DMs and their $1\sigma$ bounds are shown in the bottom three rows of Table~\ref{tab:clusts}.

Depth sampling was done for WH24-J1218 and Abell 924, since the corresponding FRB host galaxies are members of these systems. This resulted in a bimodal distribution of DM, due to the relatively higher concentration of gas compared to galaxies in these systems. Intuitively, there is a peak in the gas density closest to the center of the ICM that boosts the DM when the host is on the far side, but not on the near side. In Table~\ref{tab:clusts}, we report the medians and upper/lower $1\sigma$ percentiles for each peak separately, and in Figs.~\ref{fig:burst1_dmicm},\ref{fig:burst2_dmicm} the $s_\text{max} < 0$ (``near-side'') and $s_\text{max} > 0$ (``far-side'') samples are histogrammed with dashed and solid lines, respectively.

\begin{table}
	\hspace*{-0.75in}
	\resizebox{1.1\textwidth}{!}{%
		\begin{tabular}{@{}lcccccc@{}}
			\toprule
			\textbf{Name}  &  \textbf{Abell 576}  &  \textbf{WH24-J1218}  &  \textbf{NSCS-J1218}  &  \textbf{TullyN08}  &  \textbf{Abell 924}  \\
			\midrule
			\textbf{RA (J2000)}  &  07h21m24.10s  &  12h18m31.83s  &  12h18m42.53s  &  12h03m13.98s  &  10h07m06.39s  \\
			\textbf{Dec (J2000)}  &  55d44m20.00s  &  48d40m17.90s  &  48d43m19.09s  &  47d55m58.67s  &  35d40m41.28s  \\
			\textbf{z}  &  0.0390  &  0.1183  &  0.0448  &  0.0020  &  0.1430  \\
			\textbf{$\mathbf{M_{500}}$ [$10^{13}$~\msun{}]}  &  $\qty{147.4(13.4:12.6)}{}$  &  $\qty{6.8(4.0:2.5)}{}$  &  $\qty{3.6(2.4:1.4)}{}$  &  $\qty{6.8(2.4:1.9)}{}$  &  $\qty{21.5(12.6:7.9)}{}$  \\
			\textbf{$\mathbf{R_{500}}$ [kpc]}  &  1748  &  609  &  507  &  635  &  888  \\
			$\mathbf{c_{500}}$  &  2.31  &  2.64  &  2.77  &  2.71  &  2.48  \\
			\textbf{Source}  &  \cite{Babyk:2014}  &  \cite{Wen:Han:2024}  &  \cite{Lopes:Carvalho:2009}  &  \cite{Tully:2015}  &  \cite{Wen:Han:2024}  \\
			\textbf{FRB Sightline}  &  20231206A  &  20230703A  &  20230703A  &  20230703A  &  20230203A  \\
			\textbf{FRB Impact [$R_{500}$]}  &  2.10  &  0.70  &  0.23  &  1.13  &  0.87  \\
			\hline
			\textbf{UPP Iso [\dmu{}]} & $< 13.3$  &  $\qty{71(76:52)}{}$ / $\qty{228(148:91)}{}$  &  $\qty{938(412:273)}{}$  &  $\qty{109(48:38)}{}$  &  $\qty{35(54:26)}{}$ / $\qty{155(115:70)}{}$ \\
			\textbf{UPP Baldi [\dmu{}]}  & $< 20.7$  &  $\qty{75(74:55)}{}$ / $\qty{231(132:87)}{}$  &  $\qty{758(308:217)}{}$  &  $\qty{128(53:42)}{}$  &  $\qty{46(58:33)}{}$ / $\qty{175(116:71)}{}$ \\
			\textbf{mNFW [\dmu{}]}    &$< 80.5$  &  $\qty{74(42:42)}{}$ / $\qty{167(63:45)}{}$  &  $\qty{294(119:73)}{}$  &  $\qty{158(47:46)}{}$  &  $\qty{72(57:48)}{}$ / $\qty{188(80:63)}{}$ \\
			\bottomrule
	\end{tabular}}
    \caption{\label{tab:clusts} Physical parameters of each cluster or group in the sample, and \DM{ICM} estimates for each model under the virial convention defined in Eq~\ref{eq:virial_overdens}, normalized such that $f_g(R_v) = 1$. DM values are given in the observer frame (i.e., scaled by 1/(1+z)). When the host galaxy is a member of the group or cluster, the mean values of each peak in the \DM{ICM} distribution are given (corresponding with the host being on the ``near'' or ``far'' side of the system). For Abell 576, $1\sigma$ upper limits are shown.}
\end{table}

\subsection{DM Accounting}
\label{sec:dm_budget}


The bottom rows of Table~\ref{tab:clusts} show the median \DM{ICM} estimates for each group or cluster for each model, with $1\sigma$ uncertainties. These values are for the virial overdensity convention defined in Eq~\ref{eq:virial_overdens}, which has $\Delta \approx 103$ at $z=0$, and are normalized such that $f_\text{hot} = f_g(R_v) = 1$. For WH24-J1218 and Abell 924, we report the median of each distance-related peak (near/far) separately. The column for Abell 924 shows the results for the optical center only. For the X-ray center, the values are slightly larger: $\qty{49(74:38)}{}$ / $\qty{210(147:89)}{}$ for the UPP Iso model, $\qty{61(73:48)}{}$ / $\qty{224(146:88)}{}$ for the UPP Baldi model, and $\qty{89(59:51)}{}$ / $\qty{215(78:65)}{}$ for mNFW.

For Abell 576, the median DMs were all consistent with zero. For some mass samples within the uncertainties, the sightline intersected Abell 576 to within an $R_v$, resulting in a nonzero estimate. Table~\ref{tab:clusts} shows the $1\sigma$ upper limits of DM for each model.

The total available extragalactic DM for each FRB, after subtracting off the MW estimate, is
\begin{itemize}
	\item[] FRB 20231206A : 367.3~\dmu{}
	\item[] FRB 20230203A : 351.7~\dmu{}
	\item[] FRB 20230703A : 233.3~\dmu{}
\end{itemize}

For FRB 20231206A, the lack of ICM contribution implies that most of its DM is from the host galaxy, since at such a low redshift (0.0659) the contribution from the intervening diffuse IGM is largely negligible. Assuming the lognormal distribution for \DM{host} from \cite{Connor:2025} gives about a 5\% chance of getting this value, putting this host on the higher end of the \DM{host} distribution but broadly consistent with previous fits. Likewise, the total DM for FRB 20230203A is compatible with the range of \DM{ICM} estimates for Abell 924, on the right of Table~\ref{tab:clusts}.

The sightline to FRB 20230703A is particularly interesting, since our electron density models predict that even just one of the groups intersecting it would overwhelm the available \DM{ext}. This implies that these halos are depleted of baryons, and we can constrain their gas fractions using the extragalactic DM.

\subsection{Gas fraction constraints}
\label{sec:fg_constraint}

The \DM{ICM} estimates presented in Table~\ref{tab:clusts} all assume $f_g(R_v) = 1$, such that the total baryon fraction is equal to cosmic value $f_B$ and the stellar and cold gas mass fractions are negligible. This picture is clearly incompatible with FRB 20230703A, whose total available extragalactic $\DM{ext} = \DM{obs} - \DM{MW}$ is lower than the predicted contributions from intervening halos. For a given virial radius convention, we can set an upper limit on the gas mass fraction within that radius by applying the total DM budget. Let $\barDM{ICM,R}$ represent the DM value from Eq.~\ref{eq:dm_icm_def} integrated out to a maximum halocentric radius $R$ assuming $f_g = 1$. Also let $\DM{host}$ be the host DM, $\barDM{h}$ be $\barDM{ICM,R}$ summed over all foreground halos, and $\DM{ext} = \DM{obs} - \DM{MW}$ be the total extragalactic DM. . The maximum allowable gas fraction up to $R$ for the sightline is
\begin{equation}
	f^\text{max}_g(R) = \min \left( \frac{\DM{ext} - \DM{host}}{ \barDM{h}}, 1 \right).
	\label{eq:fmax}
\end{equation}
This is a conservative upper limit, neglecting additional contributions from the halo beyond $R$ and any other intervening matter.

To make this limit as conservative as possible, we maximize the numerator and minimize the denominator, first by letting $\DM{ext} = \DM{obs} - 0.8 \DM{MW}$ to include a 20\% uncertainty on the NE2001 model \citep{Cordes:Lazio:2002}, and assuming $\DM{host} = 0$. We minimize the denominator by adopting the $1\sigma$ lower bound on $\bar{DM}_h$. This is computed by repeatedly drawing one sample from each halo’s $10^4$ Monte-Carlo realizations (Fig.~\ref{fig:burst1_dmicm}), summing the draws to generate a distribution of $\bar{DM}_h$, and taking its 16th percentile.

Note that Eq.~\ref{eq:fmax} implicitly assumes the same $f_g$ for all three halos. Using P24's formula with the virial masses, we get $f_{g,500} = 0.28, 0.36, 0.36$ for NSCS-J1218, WH24-J1218, and TullyN08, respectively. One should thus interpret $f_g(R)^\text{max}$ as a limit on the effective gas fraction averaged over the sightline, rather than the $f_g$ of any particular halo. For a conservative comparison, we will compare our upper limits to the lowest predicted $f_g$ values among the three, while noting that the true sightline-averaged $f_g$ is likely larger.

\begin{figure}
	\centering
	\includegraphics[width=1.0\linewidth]{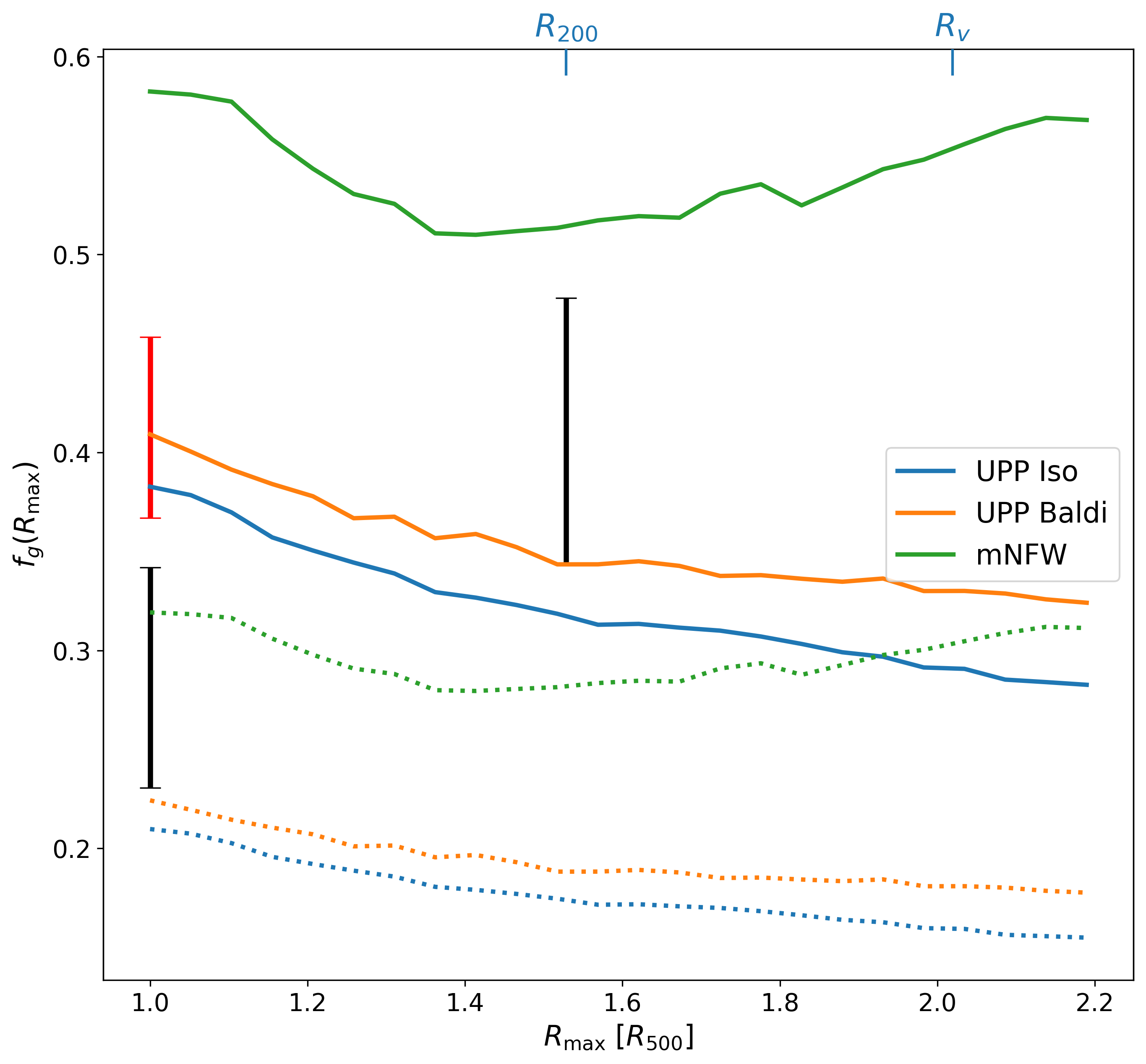}
	\caption{Upper limits on $f_g(R_\text{max})$ for each model, assuming the same $f_g$ for all three halos along the FRB 20230703A sightline. For conservative upper limits, these use the lower $1~\sigma$ bound of total $\DM{ICM}$, and only including the near-side peak of DM values for WH24-J12188 and that all three. The value at each radius ignores any contribution to $\DM{ICM}$ from outside of $R_\text{max}$, as well as any DM contributions from the host galaxy, diffuse IGMm, or large scale structure. The vertical black lines show the range of values consistent with NSCS-J1218 mass halos from the fits in P24, while the vertical red line shows the prediction from \cite{Eckert:2021}. The width of each reflects the uncertainty in the mass of NSCS-J1218. For comparison, the dotted lines show the upper limits if we subtract off $\DM{IGM} = 111$~\dmu{}, an approximate IGM contribution estimated from the $p(\DM{cosmic})$ of \cite{Hussaini:2025} and \cite{Connor:2025} (see Sec.~\ref{sec:fg_constraint}).}
	\label{fig:fhotlims}
\end{figure}

Fig.~\ref{fig:fhotlims} shows these upper bounds for each model out to various maximum radii, in units of $R_{500}$. The top x-axis marks the positions of $R_{200}$ and $R_v$ for NSCS-J1218. UPP-based models predict higher DMs here, requiring a lower $f_g$ to fit within the DM budget. The vertical black bars mark the estimates of $f_g(R_{500})$ and $f_g(R_{200})$  for NSCS-J1218  using the formulas from P24, where the heights account for the uncertainty in $M_{500}$.  Similarly, the vertical red bar shows the range of values consistent with the $f_g(R_{500})$ formula in \cite{Eckert:2021}. At $R_{500}$, the UPP-model upper limits sit above the range of values from P24 (black line), but are below the \cite{Eckert:2021} range (red line), firmly ruling it out. At $R_{200}$, the UPP models are in tension with the P24 estimates, suggesting that (at least for these systems) less baryonic matter is retained in the halo than is observed in similar-mass groups. Both X-ray derived $f_g$ estimates are consistent with the upper limits using the mNFW model.

As a point of comparison, the three dotted lines in Fig.~\ref{fig:fhotlims} show the upper limits including an approximate $\DM{IGM}$ value of $111$~\dmu{}, estimated using the relation in Eq. 4 of \cite{Hussaini:2025}, and with $f_d \approx 0.93$ as per \cite{Connor:2025}. The dotted curves use the same lower bounds on modeled DMs and upper bounds on observed DM as the solid lines. Under this less-conservative model the \cite{Eckert:2021} prediction for $f_{g,500}$ is firmly rejected, while the P24 prediction is still consistent with the mNFW model. This could suggest that, under a more realistic DM budget, the mNFW is consistent with X-ray estimates. However, we note that this still excludes any assumptions of host DM or large scale structure, so the true value is likely still much lower than shown.

There is a noticeable difference in the behavior of the UPP models and the mNFW, which can be attributed to their different behavior with $R_\text{max}$. As shown in Fig.~\ref{fig:mass_ratios}, the ratio of gas mass to total mass is gently decreasing at large radii for the UPP, while for the mNFW it increases. Since this ratio is used to normalize the total \DM{ICM} of each halo so that $f_g(R_\text{max}) = 1$, this leads to decreasing $\DM{ICM}$ for mNFW and slightly increasing for UPP. Although the $f_g$ limit is increasing at large $R_\text{max}$ for mNFW, we note that does not approach unity until almost 10 virial radii, well beyond the point where it is reasonable to assume the gas is still bound to the halo.
	
We note also that these limits are firmly incompatible with the empirical gas fraction predicted by the un-normalized UPP. From Fig.~\ref{fig:upp_fg}, the two UPP models have empirical gas fractions of $\sim 0.45 $~\textendash~$0.55$ at $10^{13}$~\msun{}, well-above the limits shown in Fig.~\ref{fig:fhotlims}. This is consistent with the expectation that lower-mass groups break with the relations observed on the higher-mass end.

\section{Discussion \& Conclusions}
\label{sec:discussion}

FRB dispersion can provide a strong upper limit on the baryon content of intervening halos, well beyond the radii typically observable in X-rays, since the FRB signal passes through the full extent of the halo. We have estimated DM contributions from foreground galaxy groups and clusters using three different ICM models, taking into account uncertainties in the masses of intervening halos, for three FRBs in the first CHIME-Outriggers host galaxy sample. For one of these (FRB 20231206A), we find the estimated DM contribution to be negligible despite the large extragalactic DM, most of its DM is from the host galaxy. While its \DM{host} is on the higher-end of what's typical, it is not a severe outlier. FRB 20230203A, embedded in the cluster Abell 924, has expected DM contributions in agreement with its total excess DM. The third sightline, to FRB 20230703A, is highly constraining on the gas content in the group mass halos that intersect it.

The total foreground DM on this last sightline, integrated out to the virial radius $R_v$ defined in Eq.~\ref{eq:virial_overdens} for each halo, is incompatible with the DM budget. This is largely due to the estimated contributions of NSCS-J1218, which alone exceeds the available DM budget, indicating that its closure radius is well beyond $R_v$. Integrating the DM to the smaller radii $R_{500}$ and $R_{200}$, we compared our $f_g$ limits to the recent eROSITA results \cite{Popesso:Biviano:2024}, and found generally good agreement out to $R_{500}$ but some tension out to $R_{200}$.

It is clear that group environments should be taken into consideration when modeling foreground DM contributions. Peculiar velocities of group / cluster members make relative distances ambiguous, complicating the process of identifying foreground galaxies when the FRB host is embedded in a cluster. Additionally, galaxies in high mass clusters are expected to have lost their CGMs to ram pressure stripping, which should reduce host DMs. Most significantly, intragroup/cluster media can contribute to the total DM in a way that is not captured by modeling individual galaxy halos. For comparison, suppose we had only considered galaxies along the line of sight to FRB 20230703A. Searching the Local Volume Sample \citep{Cook:Mazzarella:2023}, and applying the stellar-to-halo mass relation of \cite{Moster:2013}, we can identify 5 galaxies that intersect the line of sight to within one virial radius, all members of NSCS-J1218. Assuming an mNFW profile normalized to $f_{g}(R_v) = 1$, we get a total DM contribution of 277~\dmu{} from all of these. In contrast, the mNFW profile model for the IGrM predicts a total of 294~\dmu{} (as in Table~\ref{tab:clusts}) -- compatible with the galaxies-only estimate to within its error bars, but still somewhat higher. We would also have missed the likely contributions from WH24-J1218 and TullyN08 on this sightline had we not identified known galaxy groups along this sightline.

Furthermore, we have found that the uncertainties in estimates of cluster masses can lead to large uncertainties in estimated DM contributions. For host galaxies embedded in clusters, there is an additional uncertainty from the host's unknown line of sight depth. By using the NFW profile as an estimate of probabilities of galaxy positions, we can marginalize over this uncertainty, finding a largely bimodal $\DM{ICM}$ distribution. For some ICM models and impact parameters, this implies that the expectation of $\DM{ICM}$ for host galaxies on the near side is non-zero. While zero contribution from the host cluster is possible if the host galaxy is on the near-side halo boundary, this is generally improbable, and should only be assumed as a conservative lower bound.

The results presented here are somewhat dependent on the assumed mass-concentration relation from the Magneticum simulations \citep{Ragagnin:Saro:2021}. The choice of concentration parameter affects the results through a few routes. When measured quantities in the literature are in the 200 or V ($\Delta \approx 104$) convention, then the concentration parameter is used to convert to the 500 convention. The depth sampling and DM normalization steps also depend on the concentration, since both use an assumed NFW profile. Lastly, the mNFW model itself explicitly depends on concentration, in contrast to the UPP models which use a fixed $c_{500}$ value.  For comparison, we also ran the Monte Carlo analysis with concentrations derived from IllustrisTNG \citep{Sorini:Bose:2025} and the Eagle simulation \citep{Beltz-Mohrmann:2021}. With these, we see similar results to those of Magneticum. Overall, the uncertainty due to concentration appears to be a smaller effect to the mass uncertainty.

One may expect the upper limits in Fig.~\ref{fig:fhotlims} to converge to 1 at large $R$, as the integration limits approach the \emph{closure radius} of the halo. In an ideal scenario where all DM is due to the halo and the model is a perfect description, the $f_g^\text{max}(R)$ value should go to 1 as $R$ goes to the closure radius, since the modeled DM will approach the observed value. Allowing for unknown host, IGM, and MW contributions, one might still expect the upper limit to converge to some value less than 1. However, this expectation does not hold in this analysis due to the way we are normalizing the density profiles. Recall that for each halo boundary $R$, we set the model to have $M_g(R)/M(R) = f_B$ (as in Fig.~\ref{fig:mass_ratios}). In this way, we treat each $R$ as the closure radius for modeling, and compute upper limits on that assumption.

We note that we have applied some of these halo density models outside of their intended domains for several halos in the sample. The UPP was derived from X-ray and SZ observations of halos with masses above $10^{14}$~\msun, which is not the case for the three groups along the sightline of FRB 20230703A, and within $r < R_{500}$, which is less than the impact factor of the FRB 20231206A sightline. Although \cite{arXiv:2305.07080} argue that a UPP-like profile should be accurate out to $3 R_{500}$, the model used here was not fitted to that extent. Another factor to consider is that we have not included any IGM or large scale structure component to the DM. Recent work by \cite{Hussaini:2025}, \cite{2506.08932}, and \cite{Connor:2025} has shown strong evidence, from localized FRBs, that only about 25\% of baryons are expected to be bound in halos, with the rest in a diffuse IGM and in cosmic filaments. Assuming the relation from \cite{Hussaini:2025}, the mean cosmic DMs are 61, 111, and 138~\dmu{} for FRBs 20231206A, 20230703A, and 20230203A, respectively. In particular, the sightline to FRB 20230703A passes through a known supercluster MSCC-310, along a galaxy filament (Fig.~\ref{fig:0703_slice}). Our spherically-symmetric models do not capture the likely anisotropy of this particular system. Likewise, Abell 924 shows evidence of being dynamically disturbed, but we have assumed ICM models that assume a system in equilibrium.

As CHIME/FRB Outriggers continue localizing FRBs, we expect to encounter many more sightlines crossing through group-scale massive halos. Integrating a halo mass function \citep{Murray:Power:Rowbotham:2013,Tinker:2008} with the CHIME/FRB detection rate vs. z from \cite{Shin:Masui:2022}, we find that about 40\% of all FRBs that CHIME/FRB will detect will have passed through halos of mass $10^{13}$~\msun or greater. While limiting cases like FRB 20230703A may be useful for constraining gas fractions in specific systems, it is more likely that upcoming X-ray survey data will be more useful as a tool to inform foreground models for FRB research. As recent FRB results have shown large fractions of baryons outside of collapsed halos \citep{Connor:2025,2506.08932,Hussaini:2025}, better models of halos themselves will be beneficial to help map this IGM structure.

\section{Acknowledgements}

The authors wish to thank Michael McDonald and Paul Schechter for helpful discussions.

We acknowledge that CHIME and the \kkoname\ Outrigger (KKO) are built on the traditional, ancestral, and unceded territory of the Syilx Okanagan people. \kkoname\ is situated on land leased from the Imperial Metals Corporation. We are grateful to the staff of the Dominion Radio Astrophysical Observatory, which is operated by the National Research Council of Canada. CHIME operations are funded by a grant from the NSERC Alliance Program and by support from McGill University, the University of British Columbia, and the University of Toronto. CHIME was funded by a grant from the Canada Foundation for Innovation (CFI) 2012 Leading Edge Fund (Project 31170) and by contributions from the provinces of British Columbia, Qu\'ebec, and Ontario. The CHIME/FRB Project was funded by a grant from the CFI 2015 Innovation Fund (Project 33213) and by contributions from the provinces of British Columbia and Qu\'ebec, and by the Dunlap Institute for Astronomy and Astrophysics at the University of Toronto. Additional support was provided by the Canadian Institute for Advanced Research (CIFAR), the Trottier Space Institute at McGill University, and the University of British Columbia. The CHIME/FRB baseband recording system is funded in part by a CFI John R. Evans Leaders Fund award to IHS.

\allacks{}

The CHIME/FRB Outriggers program is funded by the Gordon and Betty Moore Foundation and by a National Science Foundation (NSF) grant (2008031). FRB research at MIT is supported by an NSF grant (2008031). FRB research at WVU is supported by an NSF grant (2006548, 2018490). We are grateful to Robert Kirshner for early support and encouragement in the Outriggers project.

This research has made use of the VizieR catalogue access tool \citep{Ochsenbein:2000} and SIMBAD database, both operated at CDS, Strasbourg, France. This research made use of hips2fits,\footnote{\url{https://alasky.cds.unistra.fr/hips-image-services/hips2fits}} a service provided by CDS.

\software{Astropy \citep{astropy:2013, astropy:2018, astropy:2022}, Astroquery \citep{astroquery}, numpy, scipy, matplotlib. }

%This publication makes use of data products from the Wide-field Infrared Survey Explorer, which is a joint project of the University of California, Los Angeles, and the Jet Propulsion Laboratory/California Institute of Technology, funded by the National Aeronautics and Space Administration. 

%\bibliography{clusters}
\bibliography{icm,other}

\appendix
\section{UPP Normalization}
\label{app:upp_norm}

\begin{figure}
	\centering
	\includegraphics[width=0.7\linewidth]{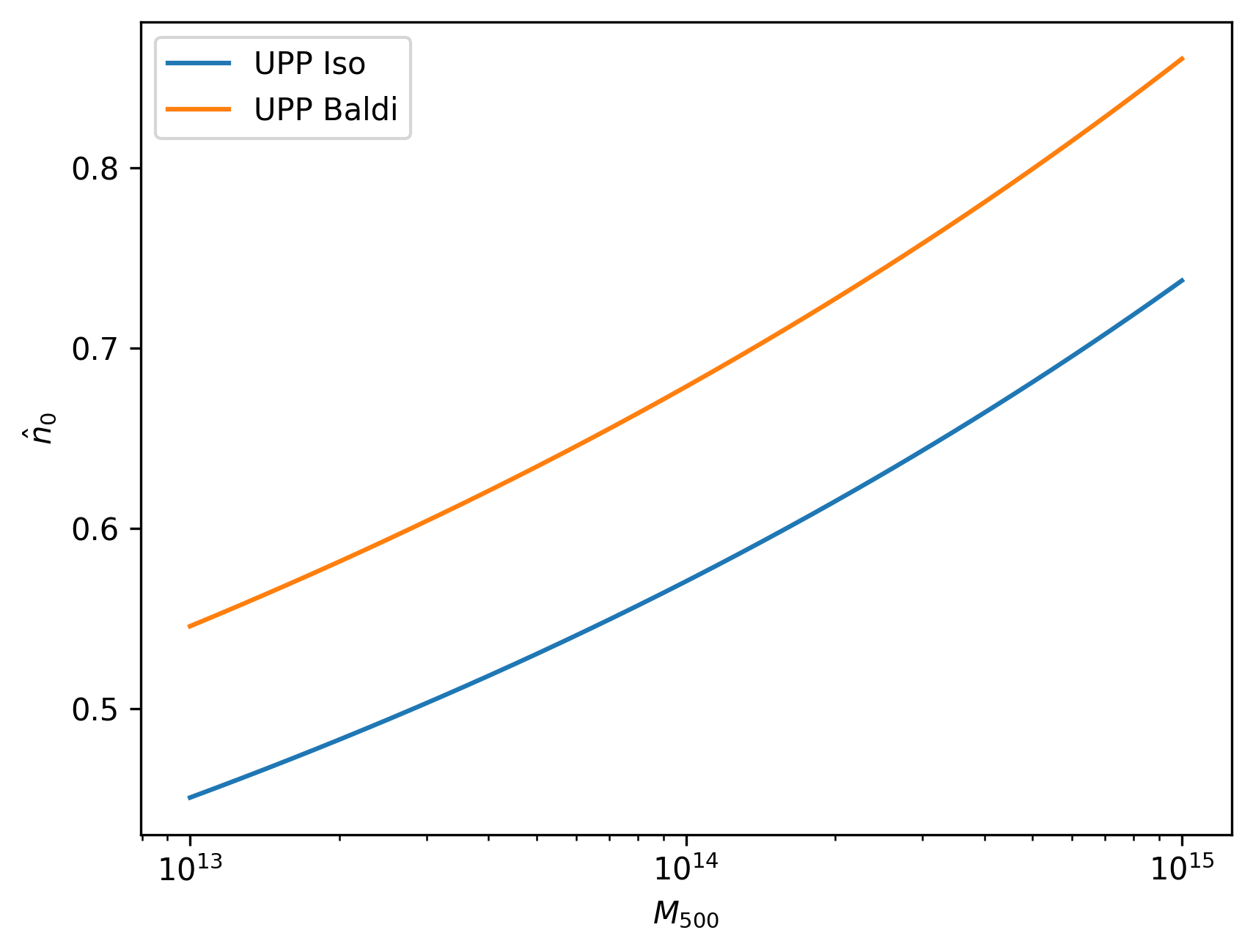}
	\caption{Numerically-integrated values of $\hat{n}_0$ for the isothermal and cool-core UPP over a range of masses.}
	\label{fig:upp_fg}
\end{figure}

The UPP, being empirically-derived, implicitly assumes a value of $f_{g,500}$. To make the gas fraction a free parameter, we need to normalize the densities such that $f_{g,500} = 1$, using the value $f_B = 0.175$ assumed in \cite{Arnaud:2010}.

The baryon mass density is related to the pressure profile via
\begin{equation}
	\rho_b = \frac{P(y)}{k_B T} = \frac{p(y) \rho_v f_B T_{500}}{T(y)},
\end{equation}
where $\rho_v = M_v / (4/3 \pi R_v^3)$ is the virial density and $p(y)$ is the dimensionless pressure profile from Eq.~\ref{eq:upp_general}. We integrate this to get $M_b$:
\begin{equation}
	M_b = R_v^3 \int\limits_0^1 4 \pi x^2 \rho_b(y) dy = 4 \pi R_v^3 \rho_v f_B T_{500} \int\limits_0^1 y^2 p(y) / T(y) dy
\end{equation}
The factor of $R_v^3$ comes from changing the integration variable to the dimensionless $y = r/R_v$. Dividing by $f_B M_v$ then yields the implicit gas fraction
\begin{equation}
	\hat{n}_{0} = 3  T_{500} \int\limits_0^1 x^2 p(x) / T(x) dx
\end{equation}

Fig.~\ref{fig:upp_fg} shows the gas fractions across a range of halo masses for both the isothermal and cool-core \cite{Baldi:Ettori:2012} temperature profiles.

\section{Virial Conventions}
\label{app:virial_conv}
The spherical collapse model motivates several definitions of the virial density, the choice of which affects how derived quantities should be interpreted. Generally, the overdensity at which a perturbation will collapse is expressed as some multiple of the critical density --- $\rho_\Delta = \Delta \rho_c(z)$. Under some assumed density profile, such as NFW, one can then solve for the total virial mass $M_\Delta$ and get the virial radius
\begin{equation}
	R_\Delta = (3 M_\Delta / 4 \pi \rho_\Delta)^{1/3}.
\end{equation}
With the NFW profile, we may also use the method in \cite{Hu:Kravtsov:2003} to convert values between different $\Delta$s.

A common convention, based on the solution for spherical collapse in a matter-dominated (Einstein-de Sitter) universe is for $\Delta \approx 200$. For galaxy clusters, it is generally more common to find virial quantities under the $\Delta = 500$ convention, as the X-ray emitting ICM is typically visible only out to $R_{500}$. Converting between these different conventions can be achieved under the assumption of a dark matter density profile, as done by \cite{Hu:Kravtsov:2003} the NFW profile.

A more realistic virial radius estimate, solving for spherical collapse in a $\Lambda$CDM universe, yields the formula \cite{Bryan:Norman:1998,Zemp:2014}:
\begin{equation}
	\Delta = (18 \pi^2 - 82 q - 39 q^2) \qquad q = \frac{\Omega_{\Lambda}}{\Omega_m a^{-3} + \Omega_\Lambda },
	\label{eq:virial_overdens}
\end{equation}
which is $\Delta \approx 103$ at z=0. Since the value of $\Delta$ here depends on redshift, we refer to this convention with the subscript ``v'' -- $R_v,M_v$.

For an NFW halo at z=0 with concentration $c_{500} = 5$, the three virial radii are related as
\begin{equation}
	R_{500} : R_{200} : R_v = 1 : 1.48 : 1.96
\end{equation}

%
%The quantity $f_g$ discussed in this paper depends on the choice of virial convention used, since $M_b$ in Eq.~\ref{eq:fb_def} comes from integrating the baryon density out to the virial radius. 
%
%With systems of galaxies, the total mass should be measurable from the velocity dispersion profile, but having an incomplete sample of member galaxies will skew this. One can apply a ``surface pressure correction'' \citep{Girardi:Giuricin:1998} to account for the effect of halo mass beyond the observed radius, under the assumption [....] For X-ray studies of galaxy clusters, the  \cite{Tully:2015a} argues that the spherical collapse model predicts that halos expand and contract multiple times as they virialize, and that the second-turnaround radius $r_{2t}$ is a more observable radius to define galaxy groups.
%

\end{document}